# The Inner Solar System's Habitability Through Time

**Anthony D. Del Genio**
*NASA Goddard Institute for Space Studies*

**David Brain**
*University of Colorado*

**Lena Noack**
*Freie Universität Berlin*

**Laura Schaefer**
*Stanford University*

## ABSTRACT

Earth, Mars, and Venus, irradiated by an evolving Sun, have had fascinating but diverging histories of habitability. Although only Earth's surface is considered to be habitable today, all three planets might have simultaneously been habitable early in their histories to microbial life. We consider how physical processes that have operated similarly or differently on these planets determined the creation and evolution of their atmospheres and surfaces over time. These include the geophysical and geochemical processes that determined the style of their interior dynamics and the presence or absence of a magnetic field; the surface-atmosphere exchange processes that acted as a source or sink for atmospheric mass and composition; the Sun-planet interactions that controlled escape of gases to space; and the atmospheric processes that interacted with these to determine climate and habitability. The divergent evolutions of the three planets provide an invaluable context for thinking about the search for life outside the Solar System.

# 1. INTRODUCTION

One of the greatest challenges in thinking about life elsewhere is that we have only one known example of a planet with life. We therefore use the criteria for life as we know it on Earth as a guide to thinking about other planets: Liquid water as a solvent for biochemical reactions; an energy source such as sunlight; chemical elements present in all life forms (C, H, N, O, P, S); and an environment suitable for life to exist and reproduce (*Hoehler*, 2007; *Cockell et al.*, 2016; see also Chapter 2 of this book).

Earth may or may not be the only planet in the Solar System that supports life now. We do know, however, that the conditions for life have continually evolved since the planets were formed. Life has existed for much of Earth's history (*Mojzsis et al.*, 1996; *Bell et al.*, 2015; *Dodd et al.*, 2017; *Tashiro et al.*, 2017; see Chapter 3), yet the physical and chemical conditions and types of life have varied over time. Mars and Venus, currently inhospitable for surface life, have their own fascinating histories that offer clues to possible habitability in earlier times.

A broader perspective on planetary habitability is obtained by considering these three neighboring planets together, at different points in their evolutions, giving us in effect 5-10 different planets to study rather than just one. This chapter focuses on external influences on all three planets over time, and how processes evolved similarly or differently on each planet due to their size and/or interactions with the Sun to cause habitability to vary from ancient to modern times. As such, this chapter lies at the intersection of Earth science, planetary science, heliophysics, and astrophysics. It builds on detailed descriptions of each planet's habitability elsewhere in this book (Chapters 1, 7, 8, 16). It also links to discussions of exoplanet properties (Chapters 15 and 20) in both directions: Solar System rocky planets through time as relatively well-explored prototypes of habitable/uninhabitable planets that inform thinking about poorly constrained exoplanets, and the larger population and diversity of exoplanets as a context for common vs. atypical aspects of our Solar System. *Baines et al.* (2013) and *Lammer et al.* (2018) provide complementary reviews of the formation and early evolution of Earth, Mars, and Venus.

The apparent link between solar composition and the building blocks of the Solar System as observed in carbonaceous chondrites (apart from volatile elements, specifically N, C and O; *Ringwood*, 1979) can also help to predict possible compositions (and hence interior structures) of rocky exoplanets, given observations of the stellar spectrum. The elemental abundances (at least of major, non-volatile elements such as Fe, Mg, and Si) should resemble those of the accretion disk and therefore of planets around other stars (*Bond et al.*, 2010). This theory is being tested with Solar System rocky planets (*Dorn et al.*, 2015; *Unterborn and Panero*, 2017). The mass-radius ratio of Mars, Earth and Venus can be explained based on solar abundances - derived from chondritic meteorites and the solar spectrum - within an error interval, whereas Mercury is an exception, having had a more complicated formation history than assumed for the other inner planets (Chapter 13). Some of the most abundant elements in the Solar spectrum (H, O, C) condense further out beyond the snowlines of the various icy compounds they form (pure ices, hydrates, clathrates). Planetesimals from these regions of the planetary disk contribute to the formation of icy moons and the enrichment of volatiles and heavy elements in gas giant atmospheres (*Johnson et al.*, 2012). The fraction of these abundant elements on the terrestrial planets depends on the extent to which they occur in minerals with high melting points as opposed to volatile ices. For this reason, O is the most abundant element in Earth's crust and the second most abundant element (after Fe) on the planet (*McDonough and Sun, 1995*).

Exoplanets on close-in orbits, for which the mass and radius are known with a sufficiently high accuracy, seem to confirm that a relation exists between element abundances in stellar spectra and planet building blocks, as observed via the densities of the planets CoRoT-7b, Kepler-10b, and Kepler-93b (*Santos et al., 2015*). Some degeneracy in the prediction of exoplanet composition cannot be avoided, though, since the migration of planets during their formation as well as the accretion history matter. This can be seen for Mercury, for which iron content is in excess of what would be predicted from the solar composition, plausibly caused by preferential removal of silicate material during a giant impact (*Benz et al., 2008*). However, additional processes near the inner edge of the protoplanetary disk may further act to fractionate silicates and metals in close-in planets. This may occur by preferential loss of lower density planetesimals (*Weidenschilling, 1978*) or photophoresis (*Wurm et al., 2013*), or by later fractionation through atmospheric escape from lithophile-rich magma ocean atmospheres (*Fegley et al., 2016*), i.e., atmospheres formed at high temperature during the planet accretion stage when giant impacts melt a significant portion of the silicates and metals that form most of the planet (*Elkins-Tanton, 2012*).

The planets of the inner Solar System probably formed from a relatively, though not completely, homogeneous protoplanetary disk of gas and dust, as indicated by isotopes of various refractory elements (see e.g. *Boss, 2004*; *Larson et al.*, 2011; *Pringle et al.*, 2013). Thus, many of the differences among them today are likely to be explained by location-dependent differences in their subsequent evolution. *Baines et al.* (2013, their Table 1) extensively document abundance and isotopic ratio differences among the terrestrial planets and the clues these provide about similarities and differences in their evolutions. We discuss several of these later in this chapter. Two major differences among the planets specifically relevant to habitability are the small size of Mars relative to Earth and Venus, which has implications for its present climate, and the timing and magnitude of water delivery to each planet (summarized by *O'Brien et al.,* 2018; see also Chapters 13 and 14).

The other major influence on the habitability of Earth, Venus, and Mars has been the evolution of the Sun. Solar luminosity has increased by ~30% since it first reached the main sequence ~4.6 Gya, due to hydrogen fusion, core contraction, and the resulting increase in core temperature. Accompanying this is a slight shift of the Sun's spectrum toward shorter wavelengths (Fig. 4 of *Claire et al.*, 2012). *Claire et al.* present parameterizations of solar flux evolution for an average G-type star, summarized in terms of the flux incident at Earth, Mars, and Venus in Figure 1. The short-wavelength solar flux (X-ray and UV) for such a star decreases over time from a very active "young Sun" to its more quiescent behavior now. There is however considerable uncertainty in the Sun's early history. *Tu et al.* (2015) show that depending on the Sun's initial rotation rate, a large range of X-ray luminosities in its first billion years is possible. The uncertain evolution of the Sun has implications for photochemistry and escape processes, as we discuss later. It is also a reminder that exoplanets we observe now are snapshots in their own evolving but uncertain climate history and that the lifetime of any planet within the habitable zone is finite (*Rushby et al.*, 2013).

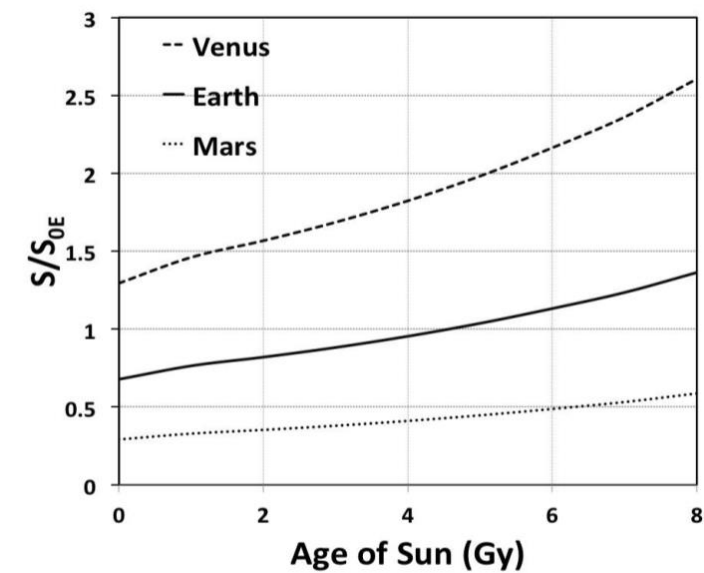

Figure 1. Temporal evolution of the broadband solar flux incident on Earth, Venus, and Mars ($S$) normalized by that incident on Earth today ($S_{0E}$), based on results of *Claire et al.* (2012). Values of $S/S_{0E}$ early in Solar System history, when Mars was probably habitable and Venus might have been, are the basis for traditional estimates of the width of the habitable zone (Chapter 20).

The net result of the initial state of each planet, the brightening of the Sun in the visible and infrared, and the decrease in X-ray and UV activity, has been that the composition, climate, and habitability of Earth, Mars, and Venus have diverged over geologic time from early in Solar System history when all three may have been habitable to the present epoch, in which Venus is clearly uninhabitable and Mars is either marginally habitable or uninhabitable (Figure 2). It is sometimes assumed in exoplanet research that atmospheric thickness and composition are a function primarily of planet mass (e.g., *Kopparapu et al.*, 2014). While this may be true in a statistical sense over the broad ranges of size and mass that differentiate the largest planets that retained their primordial thick $H_2$-He envelope and the smallest planets that lost any primordial atmosphere, it is clearly not the rule for small rocky planets with modest size differences. Consider, e.g., Venus, almost as large as Earth but with a surface pressure 92 times that of Earth, or Titan, a body smaller than Mars but with a 1.5 bar atmosphere compared to Mars' 6 mbar.

These seeming inconsistencies point to more complex histories influenced by a variety of processes: (1) Exogenous sources that deliver different amounts of volatiles to planets as a function of location within the Solar System or stochastically, determined by dynamical interactions among planets and planetesimals as the Solar System evolved; (2) Outgassing of volatiles from planet interiors and sequestration of other volatiles in the interiors as geodynamic and geochemical processes evolve toward surface-atmosphere chemical equilibrium; (3) Atmospheric loss processes, depending on the type and age of the planet's host star, its distance from that star, the presence or absence of a planetary magnetic field, and dynamic/thermodynamic/chemical processes within the atmosphere. Thus the surface habitability of any planet depends on the combined evolution of its atmosphere, surface and interior, as well as that of its star (Fig. 2). It is a great challenge to anticipate the histories of exoplanets given the meager observational constraints. Thus the more well observed terrestrial planets of the Solar System are an invaluable source of information about how these processes have played out over time for a fairly narrow range of distances in a single planetary system. We assume that the divergent evolutions of the three planets are mostly deterministic functions of their size and distance from the Sun, but stochastic sources and sinks early in their histories are likely, and bistability and bifurcations in evolution from similar initial states are possible as well (*Lenardic et al.,* 2016a).

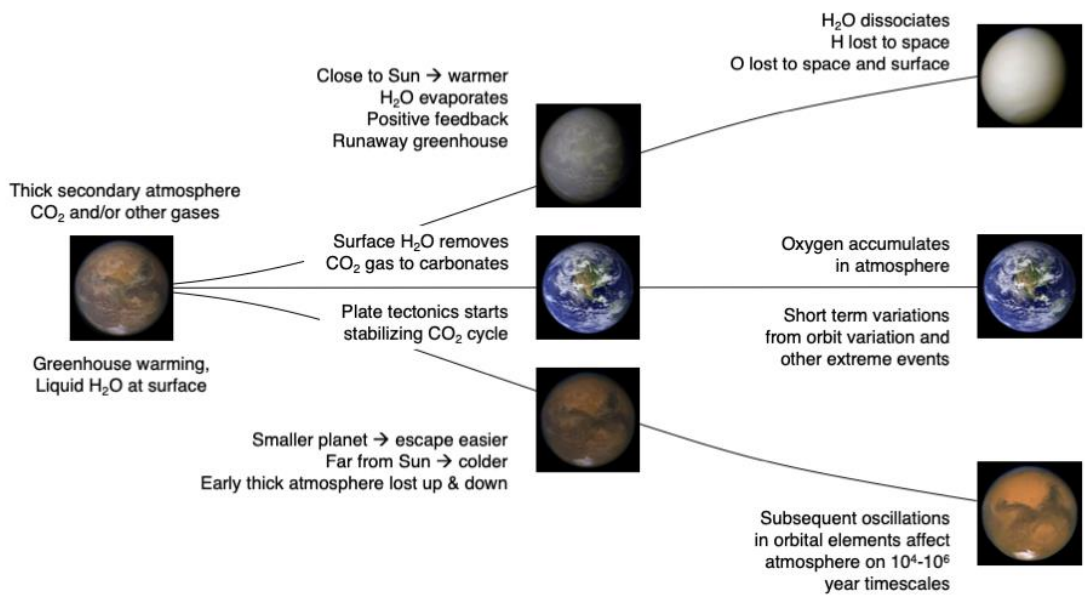

Figure 2. Venus, Earth, and Mars may have formed with similar inventories of chemical elements and could conceivably have had similar surface conditions early in Solar System history. Differences in their sizes and/or distances from the Sun, as well as the emergence of life on Earth (see Chapter 4), are the most likely causes of their subsequent divergent evolutions and explain their disparate present climates.

This chapter compares processes operating in the interiors of the three terrestrial planets (Section 2), at their surface-atmosphere interfaces (Section 3), and at their atmosphere-space interfaces (Section 4) to produce the conditions that have determined their climates and habitability through geologic time (Section 5). The chapter closes with a discussion of the perspectives these planets provide for the search for life outside our Solar System (Section 6).

## 2. PLANETARY INTERIOR PROCESSES

Planetary interior processes drive the evolution of the atmospheric composition, surface tectonics and magnetic dynamo generation, and so play multiple key roles in planetary habitability.

The interior structure and composition of the planet are set during the earliest evolution of a rocky planet, including the formation of the core that is needed for a long-term dynamo generation and the mantle chemistry. The redox state of the mantle influences the behavior of volatiles when rocks start to melt, and so the composition of gases released from the interior via volcanism that will ultimately contribute to a planet's atmosphere and climate. The interior structure and composition also affect the planet's thermal structure and rate of heat loss, which in turn factor into the tectonic mode of the planet and whether or not plate tectonics, which may help buffer climate, can occur, and how long a planet can stay geologically active. Below we describe these key characteristics of the planetary interior that are relevant to habitability.

## 2.1 Core Formation and Fractionation of Elements

Information about exoplanet interiors beyond rough density estimates for planets for which both minimum mass and radius are known is difficult to obtain, except in rare instances in which dust tails from disintegrating planets can be observed (e.g., *van Lieshout et al.*, 2014, 2016; *Bodman et al.*, 2018). Thus, knowledge about the interiors of Solar System rocky planets is crucial for constraining ideas about exoplanet formation and climate evolution. Each of the terrestrial planets of the Solar System has a central, metallic core, surrounded by a dominantly silicate mantle and crust. Earth's metallic core is the source of its magnetic field (Section 2.2), which shields the planet from harmful high-energy particles and may protect the atmosphere from certain escape processes (Section 4). Differentiation during planet formation, in which denser elements such as Fe sink to the core and lighter elements form the mantle and crust, determines the temperature, size and composition of the core. This process also influences the oxidation state of the mantle (Section 2.3) as well as the abundances of both minor (e.g., Ni, Co, C, S, etc.) and major (e.g., O, Si, Fe) elements in the primitive silicate mantle. Later silicate differentiation further fractionates elements between the mantle and crust. The size and density of Earth's inner and outer core are well known from seismic measurements and can be used to infer information about the core composition in conjunction with knowledge of siderophile element abundances in the silicate mantle. In contrast, the size, composition, and phase of the cores of Venus and Mars must be inferred through indirect observations, meteoritic abundances (for Mars), and models.

Metal and sulfide-loving (siderophile and chalcophile) trace elements will preferentially partition into the core-forming phase and be removed from the mantle (e.g., *Ringwood,* 1959; *Li and Agee,* 1996). This leads to an overall depletion of, in particular, highly siderophile elements in the observable silicate planet. Other elements may partition into the core to a greater or lesser degree depending on the conditions during separation, e.g., temperature, pressure, oxygen fugacity (a measure of the amount of free or uncombined oxygen available for chemical reactions), etc. Late stage accretion of terrestrial planets likely occurred through impacts of objects that had already undergone differentiation, suggesting that material was added to the core of the growing planet in discrete intervals. Partial re-equilibration of the cores of the accreting objects with the mantle occurs at progressively higher pressures and temperatures as the planet grows (e.g., *Rubie et al.,* 2015). Measurements of siderophile element abundances in the silicate mantle can be used to constrain the conditions of core formation for the Earth (e.g., *Rubie et al.,* 2015; *Fischer et al.,* 2015) and for Mars based on meteoritic data (e.g., *Righter et al.,* 2015).

The cores of Earth and Mars are thought to contain significant amounts of lighter elements. For Earth, a density deficit in the outer core compared to pure Fe liquid indicates that this light element(s) must make up ~10 wt% of the outer core (*Birch,* 1952), therefore indicating an abundant, in the cosmochemical sense, element must be present. Many models suggest combinations of Si, O, S, C, and H in the Earth's core (*Poirier,* 1994; *Hillgren et al.* 2000). Addition of Si and O, in particular, to the Earth's core may influence the oxidation state of the mantle. Recent models suggest up to 8.5 wt% Si and 1.6 wt% O in Earth's core (*Fischer et al.,* 2015). The inner core, however, is thought to contain only Fe, Ni alloy, causing an enrichment of the light element in the liquid outer core that may help power the geodynamo (Section 2.5). Mars' core is thought to have very large amounts of S (10-16 wt%) (*Gaillard et al.,* 2013; *Lodders and Fegley,* 1997; *Sanloup et al.,* 1999; *Wänke and Dreibus,* 1988), although a recent study of Martian meteorites suggests lower values < 5-10 wt% (*Wang and Becker,* 2017). However, the core mass fraction (0.21–0.24) and core radius (1673–1900 km) of Mars are relatively poorly known (*Rivoldini et al.,* 2011). A high proportion of S likely leads to the crystallization of a sulfide such as $Fe_3S$ rather than pure iron when the inner core nucleates. However, it remains unclear at this time if Mars has a solid inner core at the present day (*Helffrich,* 2017). NASA's Insight mission (*Banerdt and Russell*, 2017) is designed to determine these properties and more.

The core of Venus is even more poorly known. Internal structure models for Venus rely on measurements of the moment of inertia and the Love number (a set of dimensionless numbers that characterize the rigidity of a body and thus how easily its shape can change), which have proven more difficult to make than for Mars. The moment of inertia of Venus is poorly known due to the slow spin rate of the planet, the effect of drag of the dense atmosphere on orbiting spacecraft, and the short lifetime of surface landers. The potential Love number $k_2$ was derived from Magellan and Pioneer Venus spacecraft data (*Konopoliv and Yoder*, 1996). It indicates a partially or fully liquid core, but reanalysis of the data using a viscoelastic rheology (rather than fully elastic) suggests that a fully solid core cannot be ruled out (*Dumoulin et al.,* 2017). The mass and radius of Venus' core cannot be fully determined from the Love number alone, and therefore many models of Venus' internal structure use a scaled version of the Earth (e.g., *Mocquet et al.,* 2011; *Aitta,* 2012). Using a range of compositional models derived from planet formation models and different internal temperature profiles, *Dumoulin et al.* (2017) find a range of core radii of 2941-3425 km (0.48–0.57 $R_{Venus}$), with core-mantle boundary pressures of 103-127 GPa, which are all consistent with the measured total mass and radius of Venus. The assumption of a scaled version of the density of Earth's core, implying a similar abundance of light elements, must be tested by future models and measurements. Inferences about Venus' internal thermal history and core cooling and the possible existence of a magnetodynamo before the present day therefore remain speculative until additional constraints on the internal structure of Venus can be obtained.

## 2.2 Magnetic Field Development

The magnetic field may be a contributing factor to the habitability of Earth, as it shields life at the surface from harmful radiation, and may have helped to limit atmosphere erosion in the early Solar System, when the Sun emitted strong EUV radiation leading to thermal and non-thermal escape processes (*Tian et al.*, 2008); but see Section 4.4 for a more thorough discussion and some caveats. Earth is not the only body in the Solar System with an active magnetic dynamo leading to a magnetosphere. Mercury, the smallest planet in the Solar System, has a weak magnetic field of about 200 nT (*Anderson et al.,* 2011, compared to more than 50000 nT field strength for present-day Earth). In contrast, neither Venus nor Mars have an active dynamo creating a measurable magnetic field. Mars' magnetic dynamo stopped about 4.1 Gyr ago (*Morschhauser et al.,* 2018), whereas Venus shows no evidence of past dynamo activity.

A planetary magnetic dynamo needs a fluid with high electrical conductivity, convective motion, and planetary rotation (although even slow rotation like that of Venus can sustain a magnetic field if the other conditions are satisfied; *López-Morales et al.,* 2012). The conductive fluid could be, e.g., liquid iron (as for rocky planets in the Solar System), metallic hydrogen (as is the case for Jupiter) or melt in a magma ocean. Rotation leads to the Coriolis effect, which organizes flow into rolls aligned in the direction of the rotation axis. For terrestrial planets with a similar interior structure as Earth - i.e., divided into metal core, solid rock mantle and surface crust - dynamo action is driven by convection in the liquid part of the core. Different scenarios are possible, which can explain the existence or absence of a dynamo in the rocky planets of the Solar System.

A thermal dynamo refers to convection in the core driven by a strong, super-adiabatic heat flux from core into mantle. Such a dynamo would be expected in the early evolution of planetary bodies (e.g. Mars and the Moon), where after core formation and magma ocean solidification the temperatures are much higher in the metal core than in the silicate mantle, driving a strong heat flux at the core-mantle boundary. In contrast, a compositional (sometimes called chemical) dynamo refers to convection driven by density differences from heterogenous core composition - either due to freezing of the inner core (e.g. Earth), which leads to an enrichment of lighter elements above the inner core boundary, or by snow-like iron precipitate starting at the core-mantle boundary, as suggested for Ganymede and Mercury (*Hauck et al.*, 2006; *Chen et al.*, 2008; *Rückriemen et al.*, 2018).

The absence of a magnetic dynamo on present-day Venus may best be explained by a core without an inner solid part and/or by insufficient heat flux at the core-mantle boundary, since Venus' mantle may be much hotter than Earth's mantle due to the absence of a cooling mechanism such as plate tectonics in the more recent past. A fully solid core would not allow for a magnetic field today. In that case, early Venus would likely have had higher interior temperatures, such that a dynamo could have been possible but would also have ended relatively early (e.g., *Stevenson et al.,* 1983). A present day solid core on Venus would imply significantly lower light element concentration in the metal than for the Earth, since lighter elements (e.g., sulfur) strongly decrease the metal melting temperature. This is difficult to reconcile with core formation models. It would be difficult to explain the lack of a present-day dynamo on Venus for a partially molten core, except by a recent change in tectonic style (e.g., cessation of plate tectonics) or episodic lid tectonics to suppress core heat flux (e.g., *Nimmo,* 2002; *Armann and Tackley,* 2012) and halt inner core solidification. A fully molten core is consistent with the lack of a present day dynamo, but does not prohibit an early and potentially long-lived dynamo driven by thermal convection (*Stevenson et al.,* 1983; *Nimmo,* 2002; *Driscoll and Bercovici,* 2014). The nucleation of a solid inner core in the future might allow for another period of dynamo activity on Venus. The presence of magnetite below its Curie temperature in the highlands of Venus (*Starukhina and Kreslavsky,* 2002) could provide a record of a previous geodynamo, but detecting such remnant magnetism would require a surface lander.

## 2.3 Redox State of Mantle

The redox state of a terrestrial planet's mantle, which is linked to the volatile content of mantle rocks, can be inferred from trace element compositions of melt products that reached the surface (e.g., *Smythe and Brenan,* 2016). The redox state at least of Earth's early Archean uppermost mantle was similar to the present-day state (*Canil,* 1997*; Delano,* 2001*; Li and Lee,* 2004). However, the mid-ocean ridge basalt (MORB) samples dating back to 3 Gyr show a different picture of reduced mantle conditions with decreasing oxygen fugacity with depth, which would have led to reduced outgassing from melt originating from these depths (*Aulbach and Stagno,* 2016). A later oxidation of the mantle may be linked to the initiation of plate tectonics (*Mikhail and Sverjensky,* 2014). The change of reduced Hadean crust to oxidized conditions may be linked to the decline in chondritic addition of reduced species by comets (*Yang et al.,* 2014a). On the other hand, the early Archean (and possibly Hadean) already having an oxidized upper mantle (*Delano,* 2001) could reflect volatile recycling processes, even though these would be expected to have been less efficient in oxidizing Earth's mantle compared to subduction of hydrated crust. The redox state of the upper mantle may therefore not reflect the deep mantle's oxygen fugacity.

The initial redox state depends on the building blocks of Earth and on the volatiles that were able to be contained in the mantle when the last magma ocean, produced by the large impact that formed the Moon, solidified (*Hier-Majumder and Hirschmann,* 2017*; Mukhopadhyay,* 2012). During the evolution of Earth, the redox state can change, for example due to melting processes (*Parkison and Arculus,* 1999), chemical mineral changes (*Frost et al.,* 2004*; Galimov,* 2005), as well as subduction of volatiles into the mantle (*Mikhail and Sverjensky,* 2014).

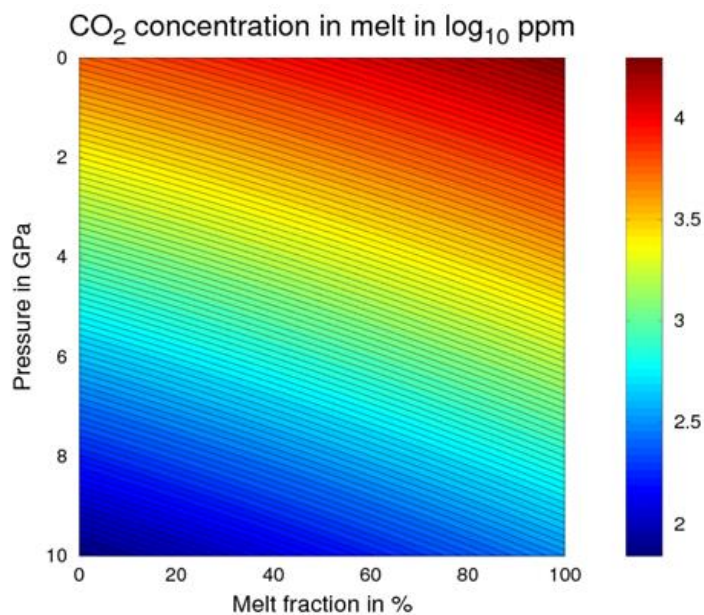

Figure 3. Amount of $CO_2$ that can be dissolved in graphite-saturated, tholeiitic melt for an oxygen fugacity at the iron-wustite buffer, based on *Holloway et al.* (1992) and *Grott et al.* (2011).

The redox state of the mantle influences the formation of carbonates in the melt and therefore has a crucial influence on the outgassing products at the surface (*Mikhail and Sverjensky,* 2014). Figure 3 shows the amount of $CO_2$ that can be dissolved in basaltic melt assuming equilibrium with graphite and an oxygen fugacity at the iron-wustite (IW) buffer (*Holloway et al.* 1992*; Grott et al.,* 2011). The melt fraction directly reflects the temperature. Melting temperatures increase strongly with increasing pressure. A melt fraction of 0 represents here the solidus melting temperature, where the first minerals of the rock assemblage begin to melt; a melt fraction of 100% represents the liquidus melting temperature, where all minerals would be molten. Increasing pressure leads to a strong reduction in $CO_2$ concentration in the melt, whereas an increase in temperature leads to higher concentrations. The oxygen fugacity is directly reflected in the amount of outgassing, as a redox state increased by one (e.g., IW+1) leads to an order of magnitude stronger outgassing. Changing oxygen fugacity in Earth's mantle at the depths where uprising melt is produced therefore strongly impacts the outgassing products. The redox state of the melt also influences the composition of the gas that is released into the atmosphere – an oxidized mantle would outgas species such as $N_2$, $CO_2$, $H_2O$, and $SO_2$, whereas from a reduced mantle $NH_3$, $CH_4$, CO, $H_2$, and $H_2S$ would be degassed.

The redox state of Mars' mantle has been inferred from geochemical analysis of SNC meteorites and has been derived to be moderately reduced (e.g., *Wadhwa,* 2001; *Herd et al,* 2002; *Grott et al.,* 2011). The oxygen fugacity of the Moon was measured in rock samples of the Apollo 12, 14 and 15 missions (*Sato et al.,* 1973). The lunar mantle seems to be more reduced in comparison to Earth, with an oxygen fugacity below the iron-wustite buffer at around IW-1. Even more reduced conditions are expected for Mercury's mantle based on high S and low FeO abundances, with an oxygen fugacity several orders of magnitude below the iron-wustite buffer (*McCubbin et al.,* 2012).

## 2.4 Heat Transport from Interior to Surface

After the accretion and magma ocean phase (Section 3.1), enormous amounts of heat are stored in the interior of planets, that are slowly, over geological timescales, transported to the surface, where the overlying atmosphere can radiate it to space. On modern Earth, the surface heat flow is estimated to be ~45-49 TW (e.g., *Davies and Davies,* 2010). This heat flux is attributed to plate tectonics, one of the processes that regulates atmospheric $CO_2$ and thus habitability (Section 3.3), helping to efficiently cool the mantle. Heat can be transported through a medium via thermal conduction, convection of material, or radiation. Inside rocky planets, only the first two mechanisms transport heat from the core to the surface. Convective currents can occur both in the liquid core and in the solid mantle, but on very different time scales. Liquid iron has a small viscosity comparable to water on the order of few mPa/s, and convection velocities are on the order of mm/s – in the solid mantle, where the viscosity of rocks is ~20 or more orders of magnitude higher, convective velocities are on the order of tens of cm/yr. Convection in the mantle only occurs if large-enough variations in the density occur (with hot, less dense material accumulating at the core-mantle boundary and cold, denser material at the top of the mantle) to trigger an instability of the system. The efficiency of convection is described by the non-dimensional Rayleigh number *Ra*, which expresses the ratio between driving and resisting factors for convection. Estimating the Rayleigh number for different planets can give a first idea of the efficiency of convective heat transport (*Breuer,* 2009). In the inner Solar System, the highest mantle $Ra$ (~$10_8$) is for Earth, closely followed by Venus (~$10_7$-$10_8$). Mars's $Ra$ is 1-2 orders of magnitude smaller (mostly due to its smaller mantle thickness $d$ ~1600 km (*Rivoldini et al.,* 2011), since $Ra \propto d_3$, and hence we expect less vigorous convection. Mercury, with a thin ~400 km mantle (*Dumberry and Rivoldini,* 2015), might not experience active mantle convection anymore. The Galilean moon Io is comparable to Earth's Moon in size but shows strong volcanic resurfacing, implying strong convection in the interior. It should be noted, though, that Io is heated constantly by tidal dissipation due to its close-in orbit around Jupiter together with tidal forces acting on Io from its neighboring moons. The surface heat flux is about 3 times higher than on present-day Earth (*Veeder et al.,* 1994), and material is efficiently extracted to the surface – which may have been similar on early Earth (*Moore and Webb,* 2013).

$Ra$ only indicates how efficient material (and hence heat) transport is over a fully convecting layer. For Archean Earth, however, mantle temperatures were higher than at present by ~150-200°C (*van Hunen and Moyen,* 2012). Higher temperatures decrease the stiffness of the mantle, and might therefore trigger stronger convection.

The other bodies of the inner Solar System do not have a crust separated into smaller plates – instead, the mantle is covered by an immobile lithosphere, which is referred to as a stagnant lid. For Venus, this may have been different during its earlier evolution – a time from which no evidence persists at the surface today as far as we know (Section 3). If plate tectonics occurs, then the surface participates in the convection cycle and leads to efficient cooling of the mantle, which may be important to maintain Earth's magnetic dynamo (Section 2.2).

While plate tectonics may have been important for the development of complex life on Earth, stagnant lid planets seem to be more abundant (given the few examples we have in the Solar System). It is therefore worthwhile to investigate the potential of stagnant lid planets with respect to surface habitability (*Noack et al.,* 2017; *Foley and Smye,* 2018; *Tosi et al.,* 2017), as they may also be abundant elsewhere in the galaxy. While planets lacking subduction and efficient resurfacing may be limited in their potential to regulate climate over long, geological timescales, planets lacking plate tectonics may still be habitable for timespans long enough for life to begin. This may have implications for the frequency with which exoplanets near the outer edge of the habitable zone can maintain habitable conditions after their early magma ocean outgassing phase.

## 2.5 Plate Tectonics and Emergence of Continents

As a result of plate tectonics, most of Earth's ancient surface rocks have been eroded and subducted over time. The rock record thus starts only ~4 Gya, leaving no direct evidence of conditions during most of the Hadean Eon, and surviving samples are found only in a few places and make up only a tiny portion of Earth's present-day surface. Oceanic crust is continuously recycled back into the mantle on time scales of tens to hundreds of Myr and is on average very young. Continental, granitic crust has a lighter composition than oceanic crust, and is more resistant against erosion and subduction. It contains the oldest outcropped rocks (for example in Greenland, Canada, western Australia and southern Africa; *Papineau,* 2010), and floats on Earth's lithospheric mantle as part of the Wilson cycle, where continental crust regularly accumulates to form supercontinents, only to break up into smaller continents spread over the entire planet's surface on geological time scales. Continental crust is formed by re-melting of previously extracted crust, for example in subduction zone settings, where water released from a subducting slab hydrates the mantle wedge and – by reducing the melting temperature of the hydrated rocks compared to dry rocks – triggers rising melt and re-melting of the base of the crust. Different ideas have been advanced for the first felsic crust (i.e., crust that has been re-melted, is less dense than primitive basaltic crust and therefore makes up the continental crust). This crust formed during the Hadean Eon under hydrated conditions, as shown by zircon minerals dating back 4.4 Gyr, which were then exposed later in Archean rocks (*Harrison,* 2009*; Arndt and Nisbet,* 2012*; Trail et al.,* 2013), as well as Archean felsic crust containing TTG (tonalite–trondhjemite– granodiorite) inclusions. These ideas include early plate tectonics, intra-plate melting, or melting at plate-like boundaries formed by uprising plumes without accompanying subduction processes (e.g. *Marschall et al.,* 2010*; Moyen and Martin,* 2012*; Rozel et al.,* 2017*; Harrison,* 2009).

Different estimates exist for the evolution of continental crustal surface coverage over time (*Hawkesworth et al.,* 2010) – ranging from an almost complete outcropping of present-day volumes of continental crust during the Archaean (*Fyfe,* 1978*; Reymer and Schubert,* 1984) to slow, increasing continental growth starting mostly at the end of the Archaean 2.5 Gya (*Taylor and McLennan,* 1995*; Breuer and Spohn,* 1995), or intermediate models with more or less steady continental growth with time (e.g., *Belousova et al.,* 2010). The rise and growth of continents, if assumed to have taken place at the end of the Archaean, has been linked to plate tectonics (*Höning et al.,* 2014) and the great oxygenation event (GOE, e.g. *Gaillard et al.,* 2011), which in turn has been linked to the increase of biovolume of single organisms (*Payne et al.,* 2009). *Rosing et al.* (2006) on the other hand suggested that energy harvested by life through photosynthesis may have played a crucial role in Earth's energy cycle and could have influenced the rise of continents. It is not clear, however, how much continental crust already existed before the GOE, and when plate tectonics initiated. Estimates go from the end of the Proterozoic Eon (~1 Gya) to as far back as the Hadean Eon (more than 4 Gya); the majority of studies suggest the geological transition of Earth's rock mineralogy observed throughout the mid-Archean to be evidence for plate tectonics initiation (see *Korenaga,* 2013, for an in-depth review on this topic).

Earth is the only rocky body in the Solar System for which evidence for currently active plate tectonics has been identified. Icy bodies such as Europa may experience a similar tectonic feature, allowing for resurfacing of the icy crust (*Kattenhorn and Prockter,* 2014), similarly to how buoyant crust is subducted on a lava lake on Earth. Venus, which today probably has surface temperatures too high to allow for active plate tectonics (*Landuyt and Bercovici,* 2009), may have experienced plate tectonics in the past, or may even be trapped in an episodic change between resurfacing and stagnant phases, determined by the interior and surface temperature evolution (*Noack et al.,* 2012*; Gillmann and Tackley,* 2014). Any signs of past plate tectonics - if it ever existed - have been erased by recent recycling of the crust in the last 300-1100 Myr (*Hansen and Lopez,* 2010). It is also not clear if Venus possesses any continental crust with a composition notably different from the lowland crust, which would be comparable to the continental-oceanic crust dichotomy that we see on Earth. The oldest crust of Venus' past, the tessera terrains, could well be felsic in composition (*Müller et al.,* 2008), but continental crust may be formed without plate tectonics, and maybe even without liquid water at the surface, and therefore even detection of felsic, continental crust on Venus does not tell us anything about its past potential habitability. Models for the early history of Venus range from Earth-like scenarios to a hellish planet from day one, with a dense runaway greenhouse atmosphere from the formation stage of the planet (Section 5).

The surface of Mars is much older in contrast to Venus and Earth, and unveils more of its early history (*Frey*, 2006). If plate tectonics or a similar crustal recycling mechanism was ever active on Mars, it must have been in the first few hundreds of Myr (within 100 Myr following *Debaille et al.*, 2009) after the magma ocean solidification stage (~20 Myr, *Bovier et al.*, 2018), since the apparently younger lowlands in the northern hemisphere are still at least 4 Gyr old (*Frey,* 2006). Also on Mars it has been suggested that part of the crust, for example in the Gale crater (*Sautter et al.,* 2015), is of continental-like composition, even though it is only 3.61 Gyr old, when plate tectonics on Mars definitely was not active. This suggests that also on Mars, continental-like crust may not have been produced by plate tectonics, but by other processes such as intrusive melt pockets leading to melting of existing crust as has been suggested for early Earth (see discussion above).

The examples of Venus and Mars suggest that the emergence of felsic, continent-like regions is not necessarily linked to plate tectonics. Continuous subduction of hydrated plates as on Earth may yield an increased production of continental crust, but this effect may be balanced by the expected increased erosion of continental crust, leading possibly to a steady-state continental crust amount exposed at Earth's surface.

# 3. SURFACE-ATMOSPHERE INTERACTIONS

## 3.1 Magma Ocean-Atmosphere Exchange: Implications for Early Water

N-body simulations suggest that giant impacts were very common during the formation of predominantly rocky planets (*Agnor et al.,* 1999; *Quintana et al.,* 2016; see also Chapter 1 of this book). Giant impacts are energetic enough to melt a significant fraction of the parent body, creating a magma ocean (*Tonks and Melosh,* 1993; *Canup*, 2008; *Nakajima and Stevenson,* 2015; *deVries et al.,* 2016). Common volatiles such as water, $CO_2$ and $N_2$, as well as more minor species such as $CH_4$, $H_2$, and $NH_3$, are soluble in silicate melts (e.g., *Holloway and Blank,* 1994; *Papale,* 1997; *Mysen et al.,* 2008; *Hirschmann et al.,* 2012; *Ardia et al.,* 2013). Therefore exchange with the atmosphere throughout the solidification process determines both the amount of volatiles that become trapped in the solidified interior, and the composition and abundance of volatiles in the atmosphere after the magma ocean solidifies (e.g., *Abe and Matsui,* 1985, 1988; *Elkins- Tanton,* 2008, *Salvador et al.,* 2017). Early oceans may result from collapse of a steam-dominated magma ocean atmosphere.

Early atmospheres for the terrestrial planets have long been thought to be dominated by water vapor and $CO_2$. $H_2O$ is more soluble in silicate melts than $CO_2$, so the abundance of water vapor in the atmosphere varies more as the magma ocean cools. In contrast, carbon partitions more readily into the metallic phase and the atmosphere. Some models suggest that for a magma ocean in contact with a metallic liquid, carbon could be pumped out of the atmosphere into the core (*Hirschmann,* 2012). The mantle oxidation state of Mars is more reduced than that of Earth (*Wadhwa,* 2008), which may favor a more reduced early atmosphere with substantial $H_2$ (*Ramirez* *et al.,* 2014; *Batalha et al.,* 2015; *Wordsworth et al.,* 2017; *Sholes et al.,* 2017), although the implications of such an atmosphere for the magma ocean and subsequent climatic evolution are still being explored.

The timing of the last giant impact, and presumably last magma ocean stage, on the Earth has been constrained by Hf-W dating of early Earth materials as well as lunar materials. These results suggest that the moon-forming impact must have occurred after the lifetime of Hf, so no earlier than ~60 Myr after the formation of Ca,Al-rich inclusions (CAIs), which are the earliest datable materials formed in the Solar System (*Kleine et al.,* 2009). Similar studies of Hf-W and Sm-Nd systematics for whole rock analyses of early Martian meteorites suggests that the Martian magma ocean cooled off within ~10-15 Myr after CAIs, with crustal formation happening no more than 20 Myr later (*Kruijer et al.,* 2017). New measurements of U-Pb ages and Hf systematics of individual zircons extracted from Martian meteorites supports an early magma ocean crystallization age and first crust formation at ~20 Myr after CAIs (*Bouvier et al.,* 2018). No direct timing constraints exist for a Venusian magma ocean since no physical samples of Venus are available for laboratory study.

Indirect evidence for a Venusian magma ocean comes from measurements of the D/H ratio in the atmosphere, which is highly enriched in D relative to the Earth and chondrites (see Chapter 16). This suggests that significant water escape could have occurred from the atmosphere of Venus during the magma ocean stage. Magma ocean models by *Gillmann et al.* (2009) and *Hamano et al.* (2013) suggest that a magma ocean stage on Venus would be significantly prolonged compared to magma oceans on the Earth or Mars, due to its closer orbital proximity to the Sun. Photolysis and escape of H from water vapor during this prolonged runaway greenhouse magma ocean stage results in significant depletion of the water reservoir and eventual solidification of the magma ocean. Thus magma ocean models suggest that it is plausible that Venus has been in its present hot, dessicated state for almost its entire lifetime. *Hamano et al.* (2013) find that planets beyond a critical distance – close to Venus' present orbital location – cool rapidly enough for the water envelope to collapse into an early ocean. Such would be the case for early Earth and Mars. If Venus was instead somewhat outside the critical distance early in its history, then enough water might have remained after the magma ocean crystallized to form an initial liquid water ocean, which then would have been lost via a later leaky moist greenhouse stage as the Sun brightened. Evidence for surface carbonates in the older regions of Venus' surface would suggest early liquid water, but the likelihood of carbonates surviving on Venus' surface is small (Section 3.2).

Crust formation on Earth and Mars was delayed towards the end of the magma ocean stage. At sufficiently high water abundance and temperature, significant rocky elements may be soluble in the water vapor atmosphere (*Fegley et al.,* 2016). Models suggest that early crust may be generated by condensation of this rocky vapor atmosphere (*Baker and Sofonio,* 2017). Other models suggest very early hydration and weathering of the earliest basaltic crust on Earth by the hot early ocean or steam atmosphere (*Abe and Matsui,* 1985; *Matsui and Abe,* 1986, 1987; *Zahnle et al.,* 1988; *Sleep et al.,* 2001). However, no record of this earliest crust on Earth exists, except perhaps in the form of ancient Jack Hills zircons (*Trail et al.,* 2011), which support the early presence of liquid water at or near the surface. The earliest crust of Mars, however, is relatively well preserved. Recent experimental work by *Cannon et al.* (2017) shows that reaction between a basaltic crust and a supercritical steam atmosphere or hot early ocean is rapid and may form thick phyllosilicate layers. Those authors simulate later crustal reworking by impacts and volcanism to show that this early clay layer matches observations of deep clay exposures and could be preserved to the present as a deep, disrupted layer in the Martian crust. The crust of Venus, in contrast, appears on the whole to be much younger (*Strom et al.,* 1994; *McKinnon et al.,* 1997) and therefore is unlikely to preserve direct signatures of the magma ocean stage.

Therefore, the case for early condensed (solid or liquid) water on Earth and Mars at the end of a magma ocean stage is relatively robust in models and seems to be well supported by the available evidence. The case for early water on Venus is harder to make if a magma ocean stage occurred, which appears difficult to avoid in planet formation models. However, the desiccation of a magma ocean atmosphere depends on the early evolution of the solar flux, which is somewhat uncertain, and radiative transfer models of thick $H_2O$-$CO_2$ atmospheres do not universally find this early dessication (see, e.g., *Kasting,* 1988, *Lammer et al.,* 2018, and discussion in Chapter 16). In fact, water loss can be quite sensitive to $CO_2$ abundance (*Wordsworth and Pierrehumbert,* 2013b). Therefore it remains uncertain whether Venus could have had early water oceans. Future characterization of the atmospheres of some of the known exoplanets slightly inside the inner edge of the traditional habitable zone (*Kane et al.*, 2014) may provide some perspective on the potential of such planets to lose vs. retain any early water. We consider both scenarios for Venus in the following sections.

## 3.2 Surface Weathering Reactions

Reactions of the atmosphere with the surface of a planet are a major sink of atmospheric gases and can therefore strongly impact its climate. The present day crusts of Mars and Venus are dominantly basaltic with small regions (<10%) of possible felsic material on both planets (*Ehlmann and Edwards,* 2014; *Gilmore et al.,* 2017). Present day Earth has a bimodal crustal distribution with felsic continents of variable age and very young basaltic oceanic crust. The growth rate of the felsic continental crust remains a highly debated topic (*Armstrong,* 1981; *Belousova et al.,* 2010; *Dhuime et al.,* 2012; *Korenaga,* 2018). Some models use ages derived from different elements to estimate the crust age distribution, while others estimate the evolution of depletion of incompatible elements from the mantle. The range of results from these two types of models prevents differentiating between the conclusion that recycling of the crust was insignificant or that it has been an important part of crustal evolution over the age of the Earth. Surface weathering rates depend on surface temperature (*Walker et al.,* 1981), as well as the composition of the surface; typically mafic minerals weather more quickly than felsic minerals (*Kump et al.,* 2000). Therefore surface-atmosphere reactions will vary with planet age based on the evolution of the crustal composition as well as climate.

The most important weathering reactions involve $CO_2$ and water (both gaseous and liquid), which can produce carbonates, as well as hydrous and oxidized minerals. The reaction of $CO_2$ with silicates to produce carbonate minerals creates a negative feedback due to the temperature dependence of the reaction and the warming greenhouse behavior of $CO_2$ in the atmosphere (*Walker et al.,* 1981; *Kump et al.,* 2000; *Sleep and Zahnle,* 2001). The formation of hydrated silicates will sequester water out of the atmosphere and ocean, which may be eventually transported to the upper mantle (see. Section 3.3). Hydration reactions may also produce $H_2$ gas as a by-product which can potentially escape and allow progressive oxidation of the planet, but may also produce greenhouse warming (e.g *Wordsworth and Pierrehumbert,* 2013a). Additional weathering reactions involve trace gases such as $SO_2$, HCl, and HF, which can produce sulfates as well as chloride and fluoride salts, and are important on both Mars and Venus, but these are expected to have less of a role in climate, so we refer the reader to the reviews by *Zolotov* (2015, 2018) for further details.

Following the magma ocean period on Earth, $CO_2$ is expected to have been very abundant in the atmosphere due to its low solubility in silicate melts, followed by a period of very rapid weathering (*Zahnle et al.,* 2010). However, rapid early drawdown of $CO_2$ on the Earth is expected to pose a problem for clement surface conditions due to the Faint Young Sun paradox unless other greenhouse gases were also present (*Sleep and Zahnle,* 2001; Section 5). Unfortunately, no geologic evidence other than remnant zircon mineral grains remain from the Hadean, making it difficult to constrain the atmospheric composition and surface conditions on the Earth for the first ~500 Myr.

On Venus, in contrast, the abundance of $CO_2$ in the atmosphere is consistent with the amount of $CO_2$ locked in sedimentary rocks on present-day Earth due to this cycle, suggesting that the majority of Venus' $CO_2$ inventory has been outgassed (*Lecuyer et al.,* 2000). At present, the carbonate-silicate cycle does not operate on Venus due to the high surface temperature and low water abundance. Rather, Venus may at present have a high enough surface temperature for mineral reactions to buffer the abundances of some atmospheric gases, although this model has been disputed in recent years (see reviews in *Gilmore et al.,* 2017, *Zolotov,* 2015, 2018) and there is at present no evidence for major gas sinks in Venus' lower atmosphere. However, the past history of surface-atmosphere interactions on Venus is not well studied, in the wet early Venus scenario.

Venus' present day surface appears to be dominated by basaltic plains, but observed tessera, which are structurally deformed materials often found in plateaus, could be composed of either mafic or felsic material (e.g., *Romeo and Turcotte,* 2008). Felsic silicates are produced by melting of other silicate materials in the presence of water (*Campbell and Taylor,* 1983), therefore the possible existence of massive felsic terrains on Venus would imply that they formed while water was still relatively abundant. If early Venus retained water, even if it resided mostly in the atmosphere, massive carbonate minerals would likely have formed. Venus' surface is sufficiently young that massive carbonates are not expected to have survived resurfacing, although they could potentially be preserved in the tessera terrain if it is truly ancient. However, carbonates are expected to be unstable on the surface of Venus today due to reaction with atmospheric sulfur compounds, and therefore any carbonates would likely have decomposed and contributed $CO_2$ to the greenhouse warming of the planet (*Gilmore et al.,* 2017; *Zolotov,* 2015, 2018).

There is ample evidence for the presence of water on early Mars, based on the abundance of clays (*Ehlmann et al.,* 2011; *Carter et al.,* 2015), the presence of valley networks (*Carr and Clow, 1981, Hynek et al.,* 2010), deltaic fan deposits (*Malin and Edgett,* 2003), and conglomerates (*Williams et al.,* 2013). However, there is evidence for only limited carbonate formation on early Mars (*Edwards and Ehlmann,* 2015; *Niles et al.,* 2013), whereas massive carbonates should be expected if Mars had an active hydrologic cycle and massive $CO_2$ atmosphere. There is a possibility that carbonate layers may be deeply layered in the crust (*Michalski and Niles,* 2010).

*Batalha et al.* (2016) suggest that the outgassing of $CO_2$ on Mars was unable to keep up with the rate of carbonate formation, causing major climatic swings from glaciated to clement in Mars' early history. In order to produce clement conditions, this model relies on $H_2$ produced by either volcanic outgassing or serpentinization of the crust to provide additional warming (*Batalha et al., 2015*).

## 3.3 Deep Volatile Recycling

Fluxes of volatiles into and out of the mantle on geologic timescales play a key role in the stability of habitable conditions at the surface by regulating greenhouse gases and replenishing volatiles from the mantle, as described in Section 3.2. Volatile fluxes out of the mantle are dictated by the rate of volcanism, whereas fluxes into the mantle depend strongly on the tectonic style of the planet as well as the weathering reactions discussed in the previous section that are responsible for removal of atmospheric gases. Mobile lid planets such as the Earth may actively transport these trapped volatiles into the mantle through subduction (e.g., *Sleep and Zahnle,* 2001; *McGovern and Schubert,* 1989), whereas stagnant lid or episodic lid planets may have more sporadic or limited transport of volatiles into the mantle (e.g., *Morschhauser et al.,* 2011).

Earth's mantle may hold ~1-10 times as much water as the surface oceans (see e.g., Table 1 of *Bounama et al.,* 2001), although more recent work discounts larger values in favor of values from 0.5-2.5 ocean masses of water (*Hirschmann and Kohlstedt,* 2012; *Korenaga et al.,* 2017), equivalent to 170-870 ppm by mass. *Hirschmann and Dasgupta* (2009) constrain the H/C ratio in the mantle to be 0.99±0.42, and the water abundance in the mantle, to be 70–570 ppm. This yields a C abundance in Earth's mantle of 90–740 ppm. The abundance of water in Venus' mantle has been estimated to be ~50 ppm at present (*Smrekar and Sotin,* 2012). Models by *Elkins-Tanton et al.* (2007) suggest that even extensive melting of the mantle should not be able to completely deplete Venus' mantle of water or $CO_2$. However, there is no estimate of the amount of $CO_2$ in Venus' mantle, and most of it is assumed to have outgassed based on analogy with Earth's total $CO_2$ inventory (*Lecuyer et al.,* 2000). Martian meteorites have been used to constrain the abundance of water in the mantle to 10-100 ppm over geologic time (*McCubbin et al.,* 2016; *Weis et al.,* 2017). The carbon content is less certain and is assumed to have been altered more significantly than water by outgassing.

Figure 4 illustrates how material exchange including volatile recycling into the mantle could occur under present conditions on Earth compared to Archaean Earth. For current Earth, the left panel shows material moving into a subduction zone, where water is released after serpentinites become unstable in the upper mantle (*Irifune et al.,* 1998; *Höning et al.,* 2014). The water flows upwards, leading to locally lower viscosities (*Hirth and Kohlstedt,* 2003) and reduced melting temperatures (*Katz et al.,* 2003) as well as more buoyant melt (*Jing and Karato,* 2009).

On early Earth (right panel of Figure 4), before plate tectonics initiated, the hydrated crustal material was probably recycled less efficiently via several different possible processes. Flat subduction under a lighter (possibly proto-continental) crust could have played a role (*van Hunen and Moyen,* 2012). Furthermore, resurfacing would have been possible by adding new crust on top of the hydrated crust, thus pushing the hydrated minerals deeper into the mantle (*Kamber et al.,* 2005; *Gorczyk and Vogt,* 2018). Sagduction, which refers to crustal material sinking into the mantle as drop-shaped material, or subduction-like behaviour on short time scales due to formation of denser minerals such as eclogite, may also have occurred (e.g., *Aoki and Takahashi,* 2004*; Rozel et al.,* 2017).

Whereas Earth has very active recycling of oceanic crust, there is no evidence for crustal recycling on Mars, which suggests that it has always been in the stagnant lid regime. Models for Mars therefore assume no volatile recycling into the mantle (e.g., *Grott et al.,* 2011; *Morschhauser et al.,* 2011; *Sandu and Kiefer,* 2012), so gases emitted into the atmosphere by volcanic outgassing remain there or on the surface until/unless they escape. In comparison, the young age of most of Venus' crust indicates at least one episode of crustal recycling. *Elkins-Tanton et al.* (2007) propose that Venus may have active recycling in the form of plume volcanism and lithospheric delamination providing a range of melting environments and source regions and a mechanism to recycle volatiles within the mantle. Using a coupled mantle-atmosphere model, *Noack et al.* (2012) find intermittent recycling of the lithosphere of Venus following periods of increased surface temperature. This would lead to return of any volatiles sequestered in the crust (perhaps massive carbonates?) back into the mantle. In a similar model, *Gillmann and Tackley* (2014) find that the $CO_2$ abundance in Venus' atmosphere did not vary significantly over time, but atmospheric water vapor is very sensitive to outgassing events, which may cause significant surface temperature fluctuations. However, these models do not allow for water oceans on Venus so the implications for volatile recycling on Venus must still be explored.

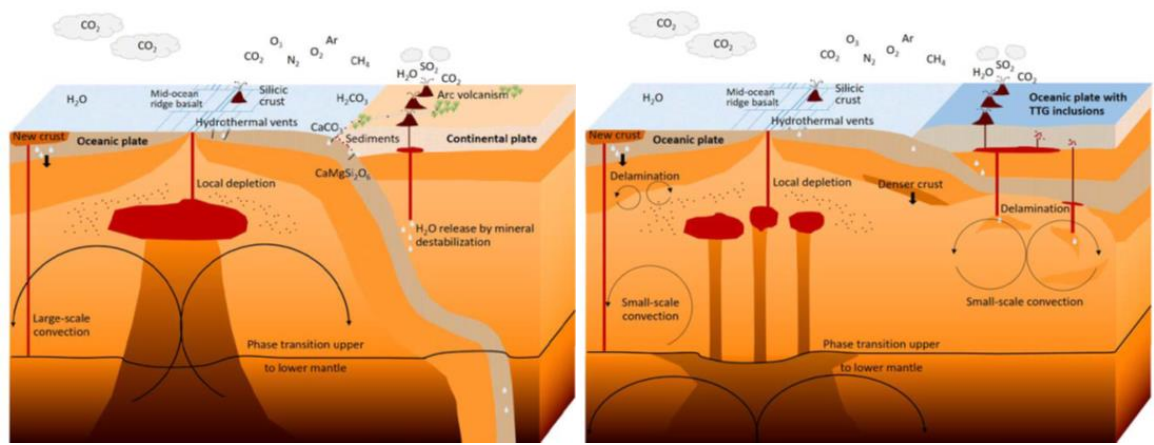

Figure 4. Sketch of Earth's present-day volatile cycles (left) including outgassing of greenhouse and trace gases, recycling of carbonates and water, and deep water release due to destabilization of serpentinites, in comparison to less efficient material recycling mechanisms that have been suggested for early Earth (right) in the absence of (present-day-like) plate tectonics and with a possible layering of convection between upper and lower mantle (as discussed by *Faccenda and DalZilio*, 2017). The depth profile of the sketch is logarithmic.

# 4. ATMOSPHERIC LOSS PROCESSES

The ability of a planet to retain an atmosphere influences whether water can be stable as a liquid at the planet's surface, and therefore strongly affects habitability. In broad terms, an atmosphere is retained when some fraction of the constituent particles is neither removed to the surface/subsurface nor removed to space. In the former case, particles can be removed through weathering (Section 3.2), adsorption, or simple surface deposition/condensation. In many cases these processes are reversible; that is, atmospheric particles removed at a planet's surface may be restored to the atmosphere if conditions near the surface change. Particles removed to space from the top of the atmosphere, however, are irreversibly lost from the system.

An individual atmospheric particle can be removed to space if it (1) is traveling upward, (2) is unlikely to collide with other atmospheric particles, and (3) has sufficient energy to escape. These three conditions must be met regardless of the mechanism that leads to their escape. The second condition requires that particles escape from the exobase region (or higher), defined as the location where the mean free path between collisions is approximately equal to the local atmospheric scale height. Regardless, a rather obvious point is that escape proceeds from the tops of planetary atmospheres. The third condition requires that a particle's velocity exceed the escape velocity $v_{esc} = (2GM/R)_{1/2}$ for the planet, where $G$ is the universal gravitational constant, $M$ the planet mass, and $R$ the planet radius. Earth (~11 km/s) and Venus (~10 km/s) have similar escape velocities due to their similar size, while smaller Mars (~5 km/s) has a smaller escape velocity. Note that more massive species must have greater kinetic energy to achieve escape velocity. An oxygen atom requires ~10 eV to escape from Venus or Earth, but a hydrogen ion requires only ~0.5 eV (these atoms require ~2 eV and ~0.1 eV, respectively, to escape from Mars). Viewed in this light, escape should be more effective at removing the atmospheres of smaller planets and lighter species. However, there are a number of potentially complicating factors, including the mechanisms for escape, whether sufficient energy or upper atmospheric particles are available to enable escape (e.g. *Zahnle and Catling*, 2017), and whether a planet's magnetic field limits escape (e.g. *Moore and Horwitz,* 2007*; Brain et al.,* 2013). How these effects combine for particular stars and planets has implications for whether a given rocky exoplanet can retain an atmosphere for a sufficiently long time to be habitable (e.g., *Airapetian et al.,* 2017; *Dong et al.,* 2017a,b*,* 2018).

Atmospheric loss to space has occurred and continues to occur on all Solar System planets via a variety of different mechanisms – all of which supply energy to upper atmospheric particles. In this section we discuss these mechanisms as they have operated at Venus, Earth, and Mars over time, and in the context of planetary surface habitability. In the three following subsections we discuss atmospheric blowoff which operated early in the histories of the terrestrial planets, contemporary escape processes (including non-thermal escape), energy- and diffusion-limited escape, and the role of planetary magnetic fields in regulating escape.

## 4.1 Atmospheric Blowoff

Shortly after their formation the terrestrial planet atmospheres likely contained substantial amounts of hydrogen and helium captured from the surrounding disk (*Baines et al.,* 2013; *Lammer et al.*, 2018). At the same time, the EUV/XUV output of the Sun, responsible for heating upper atmospheric particles, may have been much higher than it is today due to faster rotation (which is not certain - see *Tu et al*., 2015) and resultant stronger magnetic activity of the young Sun (*Ribas et al.*, 2005). The combination of low atomic mass of atmospheric species and high energy input makes it likely that early in Solar System history a significant fraction of particles in the upper atmospheres of the terrestrial worlds had enough energy to escape. Light species escaped the planets in sufficient quantities that they behaved as a fluid, or an escaping 'wind'. This process is sometimes referred to as hydrodynamic escape, and sometimes as blowoff (*Hunten*, 1973; *Watson et al.*, 1981).

Whether hydrodynamic escape occurs on a given planet depends on the temperature of atmospheric particles and the energy required to remove them. An 'escape parameter' $\lambda$ can be considered for a planet which is the ratio of gravitational potential energy to the thermal energy of the gas. This can be written as $\lambda = GMm/[kT(R+z)]$, where $m$ is the particle mass, $T$ is temperature, and $z$ is altitude. A value of $\lambda \sim 3$ is typically taken as indicating that blowoff will occur (i.e., thermal energy is at least 1/3 of gravitational potential energy). Importantly, the escaping light atmospheric particles behave as a fluid and therefore can also entrain and carry away more massive particles. Thus, species with insufficient thermal energy to escape may be removed via collisions.

The primordial $H_2$ and He atmospheres of Venus, Earth, and Mars were lost to space early in their history via hydrodynamic escape (see discussion in *Lammer et al.*, 2008). While the timing and duration of the hydrodynamic escape period is not well-constrained, there is general consensus that the $CO_2$ and $N_2$ atmospheres of these planets today are secondary, produced primarily by outgassing from the interiors. The more massive particles and declining EUV/XUV flux early in Solar System history left the atmospheres with too little thermal energy per particle to escape as a fluid. Increased attention to the topic of atmospheric loss from exoplanets has spurred a substantial amount of work on hydrodynamic escape from planets of all kinds and the role it plays in atmospheric evolution (e.g. *Yelle*, 2004; *Tian et al.*, 2005; *Owen and Jackson*, 2012; *Lammer et al.*, 2014; *Johnstone et al.*, 2015; *Fossati et al.*, 2017).

The above discussion highlights a simple but important idea when considering atmospheric evolution and habitability: a planet that has lost its atmosphere via blowoff may still develop a substantial (and habitable) atmosphere at a future time if it remains geologically active (Section 3).

## 4.2 Contemporary Escape Processes

Several different mechanisms have been responsible for removing atmospheric particles of the terrestrial planets to space over time (Table 1). The most straightforward of these is thermal escape, whereby the portion of the velocity distribution of upper atmospheric particles in excess of the escape velocity is able to leave the planet. Hydrodynamic escape is an extreme form of thermal escape, where the fraction of the distribution above the escape velocity is large (achieved as the escape velocity approaches the average thermal velocity of the distribution). Thermal escape is negligible for species heavier than hydrogen on all of the terrestrial planets today. However, it is the dominant mechanism by which hydrogen is removed from the Martian atmosphere (*Chassefière and Leblanc*, 2004), and is significant for Earth as well (~1/3 of hydrogen loss is via thermal escape). Temperatures near the Venus exobase are too low, because of cooling via $CO_2$ thermal emission, for thermal escape to be significant (*Lammer et al.*, 2008).

Energy to drive escape also comes from photochemical reactions in the upper atmosphere. Molecular species (e.g. $O_2$, $N_2$, CO) can be "photoionized" by interaction with UV light, producing an ion and free electron (e.g., $h\nu + O_2 \rightarrow O_{2+} + e$). They then dissociate when they recombine, giving the resulting atoms sufficient velocity to escape. Photochemical reactions are the dominant mechanism by which oxygen is removed from the atmosphere of Mars (e.g., *Cravens et al.*, 2016; *Lillis et al.*, 2017). These same reactions at Venus and Earth, however, do not produce atoms with velocities above the escape velocity.

Atmospheric particles are removed as ions as well, and electric fields provide the energy to enable them to escape. Ions escape from Earth's atmosphere near the poles, where the planet's global magnetic field lines are oriented vertically and are open to the solar wind. In these locations vertical electron pressure gradients are the source of electric fields that accelerate ions ($O_+$, $H_+$, etc.) upward.

Two additional sources of electric field are relevant at the unmagnetized planets Venus and Mars. First, the flowing solar wind and its embedded interplanetary magnetic field (IMF) combine to create a motional ($v \times B$) electric field that accelerates planetary ions above the exobase region. The ions are picked up by the solar wind flow and carried downstream from the planet. This electric field is important at Venus and Mars because the flowing solar wind penetrates much closer to the planet than at magnetized Earth, so that exospheric ions actually experience this field (e.g., *Luhmann and Schwingenschuh,* 1990; *Brain et al.,* 2016).

Second, interplanetary magnetic field lines drape around the conducting ionospheres of Venus and Mars, and are carried "up and over" the planetary obstacle by the flowing solar wind. However, the solar wind far from the planet is able to flow much more quickly than the solar wind that encounters the ionosphere, so that the draped magnetic field lines being carried by the solar wind can be highly curved or bent close to the planet. The bent magnetic field lines will make current flow near the location of the bend, and a corresponding Hall ($J \times B$) electric field will accelerate planetary ions downstream. In some sense, the curved magnetic field lines "scoop and slingshot" planetary ions away. Again, this process is relevant for Venus and Mars because the interplanetary magnetic field lines penetrate closer to the planet than at Earth, and therefore encounter regions where significant planetary ions are available to be accelerated (e.g., *Halekas et al.*, 2017).

We note that all three electric field terms can contribute to atmospheric escape at a planet, and even for a single planetary ion. For example, pressure gradients and/or $J \times B$ electric fields can accelerate a particle at low altitudes where the contribution from pickup is negligible, and $v \times B$ can provide additional acceleration at high altitudes where the other terms are small.

A final process of sputtering is considered for the unmagnetized planets Venus and Mars, though it has yet to be unambiguously observed (*Leblanc et al.,* 2018). Sputtering occurs when particles near the exobase region are removed via collisions with incident energetic particles – usually recently accelerated planetary ions. The contribution of sputtering to atmospheric escape at Venus and Mars today is likely small, but may have been larger in the past. Sputtering is also likely to be the dominant removal mechanism for species such as argon, that are massive (so thermal escape is unlikely) and do not react chemically (so photochemical escape cannot occur).

When considering the role that atmospheric escape plays in habitability it is important to consider the influence that the contemporary loss processes discussed above have on both overall thickness of an atmosphere (i.e. surface pressure or total atmospheric abundance) and its composition (e.g. reducing vs. oxidizing atmosphere; partial pressure of water, etc.). It is thus important to keep in mind that many distinct mechanisms remove atmospheric particles to space today, and that the different mechanisms are important for different species.

Argon isotopic ratios ($^{38}Ar/^{36}Ar$) measured in Mars' upper atmosphere indicate that Mars has lost most of its early secondary atmosphere (*Jakosky et al.*, 2017), consistent with geologic evidence for ancient clement surface temperatures and liquid water (Section 5). Extrapolation back in time of the present-day escape rate estimates for each process, accounting for a more active early Sun, suggests that escape to space has been a dominant process in the evolution of the Martian atmosphere, and that 0.5 bars (or more) of $CO_2$ was lost to space (*Jakosky et al.*, 2018). The presence of a large $CO_2$ atmosphere poses problems for some photochemical models, however, which suggest that additional greenhouse gases are required to keep such an atmosphere stable (*Zahnle et al.*, 2008), and even then large loss rates would result (*Tian et al.*, 2009). *Tian et al.* (2009) suggest that the high early escape rates created a thin, cold early Noachian Mars atmosphere and that the subsequent decrease in escape allowed outgassing to thicken the $CO_2$ atmosphere, explaining the timing of the valley networks. *Kurokawa et al.* (2018), who assume a cooler exobase temperature, estimate at least a 0.5 bar atmosphere at this time. Other escape processes such as impacts (e.g. *Brain and Jakosky*, 1998; *de Niem et al.*, 2012) or sequestration of $CO_2$ in the subsurface (e.g. *Dobrovolskis and Ingersoll*, 1975; *Stewart and Nimmo*, 2002; *Zent et al.*, 1987) may have also been important.

There is still considerable disagreement over the present-day mass loss rate at Earth (*Yau et al.*, 1988; *Seki et al., 2001*; *Slapak et al.*, 2017), as well as the timing and total amount of mass loss. Atmospheric escape has occurred in large enough quantities, however, to leave xenon mass fractionated. This has presented a puzzle for Earth, since it is assumed that the xenon was removed via hydrodynamic escape, yet other noble gases such as krypton and argon are not as strongly fractionated as xenon despite being less massive (and therefore more easily removed, one might guess) (*Ozima and Podosek*, 1983). One recent solution that has been proposed involves the escape of xenon in ionized form, since xenon is more easily ionized than most atmospheric species (*Zahnle et al.*, 2019), with the consequence that substantial atmospheric escape persisted for long periods of time as opposed to subsiding early in Earth's history. Still, models suggest that escape would have been robust early on; *Airapetian and Usmanov* (2016) estimate from a 3D model that the solar wind at 1 AU 3.95 Gya was twice as fast, 50 times denser, and twice as hot as today, leading to early mass loss rates 1-2 orders of magnitude greater than today. Compared to Mars, however, Earth's atmospheric evolution has been much more complex due to tectonic activity (Section 2) and the resulting surface-interior exchange of gases over time (Section 3). In fact, whether Earth's atmosphere is thicker or thinner today than in its past is not agreed upon (Section 6).

Even less is known about Venus' atmospheric evolution: It is marginally close enough to the Sun that it may have experienced an extended hot magma ocean phase and catastrophic Jeans and hydrodynamic loss of its early atmosphere (Sections 3, 5). A subsequent volcanic period (Sections 2, 3) is likely the proximate source of its current thick atmosphere. Venus' current atmospheric mass loss appears to be due primarily to nonthermal solar wind-induced escape (*Barabash et al.,* 2007a,b; *Brain et al.,* 2016).

### 4.3   Upper Limits on Escape: Energy-Limited vs. Diffusion-Limited

The processes described in Section 4.2 all result in the loss of atmospheric particles to space, and the difference between them on some level comes down to the differing physics of the mechanisms. When considering the flux of particles of a given species from an atmosphere, however, it is useful to ask whether the planet is in an energy-limited or diffusion-limited regime.

In an energy-limited situation, there is insufficient energy arriving to the atmosphere (in the form of solar EUV/X-ray and solar wind) to remove all particles of a given species near the exobase. Thus, an increase in energy input to the atmosphere (from a solar storm, for example, or over the course of a solar cycle) increases atmospheric escape rates. To first order, one can expect that the major atmospheric species of Earth, Mars, and Venus and their dissociation products are energy-limited because their abundance is so high at the top of the atmosphere that it is difficult to supply enough energy to remove them all.

In a diffusion-limited situation, there is ample energy to drive escape of a given species, so that a particle transported from below to the exobase region will be quickly removed from the atmosphere. In this limit the escape rates for specific escape mechanisms cease to be important; instead the rate of supply of species to the upper atmosphere controls the escape rate. One expects trace atmospheric gases to be diffusion-limited, applying similar logic to the paragraph above. These gases, such as hydrogen in a terrestrial planet atmosphere, can be critically important for surface habitability.  For example, as insolation increases on a planet with surface liquid water, moist convective storms in the troposphere may deepen and increasingly inject water into the stratosphere, where photodissociation can occur and the resulting H atoms can escape to space. The extent to which this occurs depends on the strength of the tropopause cold trap, which depends on the partial pressure of non-condensing gases in the atmosphere (*Wordsworth and Pierrehumbert*, 2014; *Kleinböhl et al.,* 2018).  Thus diffusion-limited escape of H can be tied to tropospheric moist convective storms (or, in the case of Mars, dust storms [*Chaffin et al.*, 2017]) on the terrestrial planets. The same may not be true, however, of planets orbiting cool stars, where radiatively driven large-scale ascent supplies stratospheric water instead (*Fujii et al.,* 2017).

*Kasting et al.* (1993) estimate that diffusion-limited escape would remove an Earth ocean's worth of water over the age of the Solar System once the molar mixing ratio reached $\sim 3 \times 10^{-3}$. This is the basis for the "moist greenhouse" definition of the inner edge of the habitable zone (see Chapter 20).  Modern Earth is far from this state, but it may be Earth's fate in ~1-2 Gyr, as solar luminosity continues to increase (Fig. 1; see also *Wolf and Toon*, 2015).  Whether the same process played an important role at some point in Venus' past, given its present absence of water, depends on whether it ever had a surface ocean (see Section 5). *Zahnle et al.* (2008) argue that hydrogen escape on modern Mars is at least close to the diffusion-limited rate.

## 4.4   Role of a Planetary Magnetic Field

Earth today differs dramatically from Venus and Mars in that it has a strong magnetic field. This implies that its upper atmosphere interacts less directly with its local plasma environment than the currently unmagnetized Mars and Venus (*Brain et al.*, 2016). A debate has arisen in the past decade about the importance of planetary magnetic fields in determining atmospheric escape rates (*Moore and Horwitz*, 2007; *Strangeway et al.*, 2010; *Brain et al.*, 2013). The resolution of the debate will have large consequences for our understanding of habitability of both the terrestrial planets in our own Solar System and likely candidates for habitable exoplanets.

The debate arose when it was noticed that the escape rates of ions from Venus, Earth, and Mars were the same, to within 1-2 orders of magnitude, despite the fact that Earth possesses a global dynamo magnetic field and Venus and Mars do not (*Strangeway et al.,* 2010). It was proposed that, contrary to conventional wisdom, magnetic fields do not substantially reduce atmospheric escape – that Earth's magnetic field does not actually shield the atmosphere from being stripped by the solar wind. Instead, energy from the solar wind (i.e. Poynting flux) is transferred to the upper atmosphere along magnetic field lines, and is concentrated at the poles in magnetic cusp regions. The energy then drives ion outflow from the cusps, resulting in escape rates comparable to the situation if Earth had no global magnetic field.

At present the debate is unresolved, though at least four lines of argument suggest that a planetary magnetic field has at least some effect on ion escape rates. First, measurements of ion escape from Earth and Mars as a solar storm passed both objects suggest that Mars responded much more strongly to the event than Earth (*Wei et al.,* 2012). However, solar storms evolve as they propagate away from the Sun and vary laterally, so that Earth and Mars likely did not experience identical driving conditions. Second, both models and *in situ* spacecraft observations show that the ion escape rate from Mars varies as it rotates. Mars has strong crustal magnetic fields (*Acuña et al.,* 1999) that form "mini-magnetospheres" in some geographic locations. Variation in ion escape as Mars rotates suggests that the crustal magnetic fields reduce the escape of atmospheric particles overall, but that the effect is different when the strong field is on the dayside vs. the nightside (e.g. *Ma et al.*, 2014; *Fang et al.*, 2015; *Ramstad et al.,* 2017). However, the magnitude of this effect ranges from ~10% to a factor of 2.5, depending on the analysis. Third, recent work employing theory and simple models suggests that a global magnetic field does influence escape rates (*Blackman and Tarduno*, 2018; *Gunell et al.*, 2018), though in one case a magnetic field actually increases escape (*Gunell et al.*, 2018). However, the arguments are still mostly conceptual at this stage, and wait to be followed up with rigorous models. Fourth, recent global plasma models show that ion escape rates vary as the strength of a global magnetic field is increased (*Sakai et al.*, 2018; *Egan et al.*, 2019). All of the above assume that the magnetic field extends above the atmosphere; if the magnetic field is weak, or the atmosphere is substantially inflated (*Lichtenegger et al.*, 2010; *Lammer et al.*, 2018), then the magnetic field may play a minimal role.

We note two additional caveats that make obtaining an answer challenging. First, Venus, Earth, and Mars are different sizes and orbit at different distances from the Sun. Thus one would not necessarily expect the three planets to have comparable escape rates even if they all had magnetic fields (or not) and identical atmospheres. Mars might offer a preferable alternative control experiment since it has both magnetized and unmagnetized regions on a planet of a single size, atmospheric composition, and distance from the Sun. Second, the discussion here has focused solely on ion escape, whereas atmospheric escape results from processes that affect neutral particles as well. These neutral processes are, to first order, insensitive to magnetic fields, suggesting that if the escape of a given species from a planet is dominated by one of these mechanisms then the presence or absence of a global magnetic field should not much matter.

From the perspective of planetary habitability, it would be exciting to have an answer to this question. Then we would better understand whether Earth's magnetic field played an important role in making it habitable, in contrast to Venus and Mars. And, as it becomes possible to detect exoplanetary magnetic fields, we would have an additional lever arm to use in identifying which planets in their habitable zone are most likely to be able to support life.

## 5. CLIMATE EVOLUTION

### 5.1 Planetary Energy Balance and Climate

The terrestrial planets probably formed from similar parts of the solar nebula (Section 1). Thus the diverging habitability of Earth, Mars, and Venus over time are due to differences in (a) incident solar flux, (b) partitioning in the abundances of molecules between the atmosphere, ocean and interior due to chemical and internal processes (Sections 2, 3), (c) escape of gases to space (Section 4), (d) how these processes combined to determine how much of the incident sunlight is absorbed rather than reflected back to space by each planet, and (e) how the temperature at the surface relates to that at higher altitudes where a planet with an atmosphere emits heat to space.

Mars and Venus have never been illuminated as strongly and weakly, respectively, as Earth has at any time, and the solar heating of Earth has changed by ~30% over its history (Fig. 1). Yet Earth has been habitable for most of this time (see Chapters 1, 3), Mars had an early habitable phase (see Chapter 7), and even ancient Venus might have once been habitable (see Chapter 16). Thus incident stellar flux by itself (one of the only pieces of information currently available to astronomers for their initial assessments of exoplanet climates) is very limited as a guide to whether a given planet is habitable, and if so, what its climate might be like. To understand habitability, we need to consider the series of processes by which incident stellar flux ultimately determines the surface temperature of a planet.

In equilibrium, a planet's absorbed solar flux equals its emitted thermal flux. This balance can be expressed as $[S_o(1-A)/4d^2] = \sigma T_{e4}$, where $S_o$ is the solar constant, $A$ the planetary (Bond) albedo, $d$ the planet-Sun distance in AU, $T_e$ the equilibrium temperature (i.e., the blackbody temperature that is consistent with the planet's thermal emission to space), and $\sigma$ the Stefan-Boltzmann constant (see Chapter 20). All of these quantities are measured for modern Earth, Mars, and Venus, but $A$ is not known for these planets at different times in their past and therefore must be modeled. (The same is true for exoplanets, for which astronomers just assume a value of $A$ to estimate $T_e$ in the absence of reflected light phase curves.) $T_e$ is not the quantity of interest for habitability, however, because a planet with an atmosphere containing absorbing gases emits most of its heat to space from higher altitudes within the atmosphere rather than from the surface. The surface temperature $T_s > T_e$ by an amount that depends on the extent to which surface emitted heat is absorbed by the atmosphere and re-radiated downward. This can be expressed as $\sigma T_{s^4} = \sigma T_{e^4} + G(\tau)$, where $G$, the greenhouse effect of the atmosphere, increases with $\tau$, its thermal infrared optical thickness (*Raval and Ramanathan*, 1989). An optically thin atmosphere ($\tau \lesssim 1$) may be in radiative equilibrium, with $T_s$ determined only by radiative processes. In an optically thick atmosphere ($\tau >> 1$), convection sets in to moderate $T_s$, producing radiative-convective equilibrium (*Manabe and Strickler*, 1964). Thus to know the surface temperature, the incident stellar flux, the fraction of that flux reflected to space by the planet, and the composition and pressure of the atmosphere (which determine how opaque it is to thermal radiation) must all be constrained.

The roles of individual greenhouse gases depend on their chemical stability, temperature dependence, and wavelengths of their absorption bands. Climate is controlled by greenhouse gases that do not condense at the temperatures typically found in a given atmosphere, regardless of the relative warming by each gas. Thus, Earth's temperature is regulated by $CO_2$ (which does not condense at temperatures encountered on Earth) even though $H_2O$ contributes more warming. This occurs because atmospheric $H_2O$ vapor concentration increases strongly with temperature via the Clausius- Clapeyron equation (modulated by weather-related changes in relative humidity). If $CO_2$ and its greenhouse effect were removed, the climate would cool, causing water to condense and the $H_2O$ vapor concentration to decrease, producing a weaker greenhouse effect and thus further cooling, leading to further reduction in $H_2O$ vapor, further cooling, and so on. In other words, $H_2O$ vapor cannot act on its own as a greenhouse gas on Earth - it only affects climate as a feedback that amplifies $CO_2$ warming, which can be considered an external forcing at temperatures at which it does not condense (*Lacis et al.*, 2010). For planets warmer than Earth, with atmospheres where $H_2O$ condensation is unlikely, then $H_2O$ dominates the greenhouse effect. For planets colder than Earth, gases that can avoid condensation at temperatures that would freeze $H_2O$ (e.g., $CO_2$, $CH_4$, $H_2$) become more important than $H_2O$.

Bond albedo depends on the clear atmosphere, condensate clouds, other atmospheric particulates ("aerosols" or "hazes"), and the surface (see discussion in *Del Genio,* 2013). Rayleigh scattering by gases is a minor contributor except for cold planets with very thick atmospheres. Aerosols can be a minor (Earth) or major (Venus) contributor to albedo. Condensates account for ~⅔ of Earth's Bond albedo (*Stephens et al*., 2012), but their role varies from one planet to another depending on the water (and other condensable) abundance and circulation regime. Surface albedo can be a major or minor contributor to Bond albedo depending on masking by the atmosphere. Earth's surface albedo contributes only ~12% of its total ~0.3 Bond albedo (*Donohoe and Battisti*, 2011). On Venus, it contributes virtually nothing because of a planet-wide thick sulfuric acid haze. On cold planets or planets with thin atmospheres (e.g., Mars), the surface is important to the radiative balance. General circulation model (GCM) simulations indicate that habitable exoplanets orbiting G stars that are more strongly or more weakly irradiated than modern Earth will more likely than not have higher Bond albedos because of the enhanced contributions of clouds or sea ice/snow, respectively, while most habitable planets orbiting M stars will have lower Bond albedos than modern Earth because of the enhanced absorption of the mostly near-IR incident radiation by water vapor and sea ice (*Shields et al.*, 2013; *Del Genio et al.*, 2019).

## 5.2 Implications for the Evolution of Climate on Earth, Mars, and Venus

### 5.2.1 The Faint Young Sun problem (or opportunity)

Earth's surface sustained liquid water and life even in the early Archean Eon (3.8-2.5 Gya), when the Sun was ~25% dimmer than today (Chapter 1). *Sagan and Mullen* (1972) noted the paradox between the faint Sun and the origin of life, estimating that a modern Earth-like Archean atmosphere would have a surface temperature below freezing, and thus a larger greenhouse effect was needed to maintain liquid water on the surface. *Walker et al.* (1981) proposed the carbonate-silicate cycle feedback as a reason for elevated Archean $CO_2$ to explain this (Section 3). This is also assumed to be relevant to the habitability of ancient Mars and to define the outer edge of the habitable zone (see Chapter 20). Models often just assume that more weakly irradiated planets have higher $CO_2$, but some represent the carbonate-silicate cycle control on $CO_2$ interactively (*Menou*, 2015; *Batalha et al.*, 2016; *Haqq-Misra et al.*, 2016; *Charnay et al.*, 2017; *Krissansen-Totton* et al., 2018). The ancient Earth and Mars climate problems raise the question of what $CO_2$ levels might plausibly be expected on weakly irradiated exoplanets. This has led several groups to explore other controls on the weathering sink, such as planet rotation rate (*Jansen et al.*, 2019), and on the outgassing source, such as planet size and tectonic regime (*Noack et al.,* 2017; *Rushby et al.*, 2018; see Section 2.4). It has also led to proposals to derive statistical constraints on the feedback by observationally constraining $CO_2$ abundances on a large number of exoplanets with a range of incident stellar fluxes (*Bean et al.*, 2017).

As a result, it is fair to say that the faint young Sun paradox no longer exists for Earth, although it cannot yet be considered solved because the observational evidence on our own planet is inadequate to constrain whether Archean climate was cool or very warm (*Feulner*, 2012). Coupled climate - carbon cycle models predict a high (relative to modern Earth) ~100 mb of $CO_2$ (*Charnay et al.*, 2017; *Krissansen-Totton et al.*, 2018). Observational inferences of Archean $CO_2$ vary by orders of magnitude (*Olson et al.,* 2018), from ~0.9 mb (*Rosing et al.*, 2010) to hundreds of mb (*Kanzaki and Murakami*, 2015). The absence of an $^{14}N/^{15}N$ isotopic anomaly in Earth's atmosphere (e.g., *Mandt et al.,* 2015) suggests that its exobase was never high enough for significant escape of N to occur. This in turn implies that sufficient $CO_2$, much higher than the *Rosing et al.* (2010) estimate, was present on early Earth to radiatively cool the upper atmosphere and keep the exobase altitude sufficiently low to protect it from an enhanced early EUV flux (*Lammer et al.*, 2018).

$CH_4$ is also poorly constrained; the Archean pre-dates the rise of $O_2$ and thus $CH_4$ was more chemically stable and probably much more abundant than in today's oxidizing atmosphere. Plausible $CH_4$ concentrations (*Olson et al.*, 2018) add ~10-15°C to estimated global temperatures (*Charnay et al.*, 2013; *Wolf and Toon*, 2013), but the actual role of $CH_4$ at this time depends on the strength of biological fluxes. Depending on how much more vigorous tectonics was on early Earth and thus how efficient the carbonate-silicate cycle was (Section 3), $CH_4$ (or another greenhouse gas) might have been needed not only in the early Archean but even in the prior Hadean Eon to maintain warm temperatures after Earth's magma ocean phase (*Sleep and Zahnle*, 2001).

One possibility is outgassed H₂ (*Wordsworth and Pierrehumbert,* 2013a), depending on its early rate of escape (Section 4). Another is photochemical N₂O produced from energetic particles associated with strong coronal mass ejections (*Airapetian et al.*, 2016), but whether this process could affect the troposphere enough to matter for the climate is not yet known.

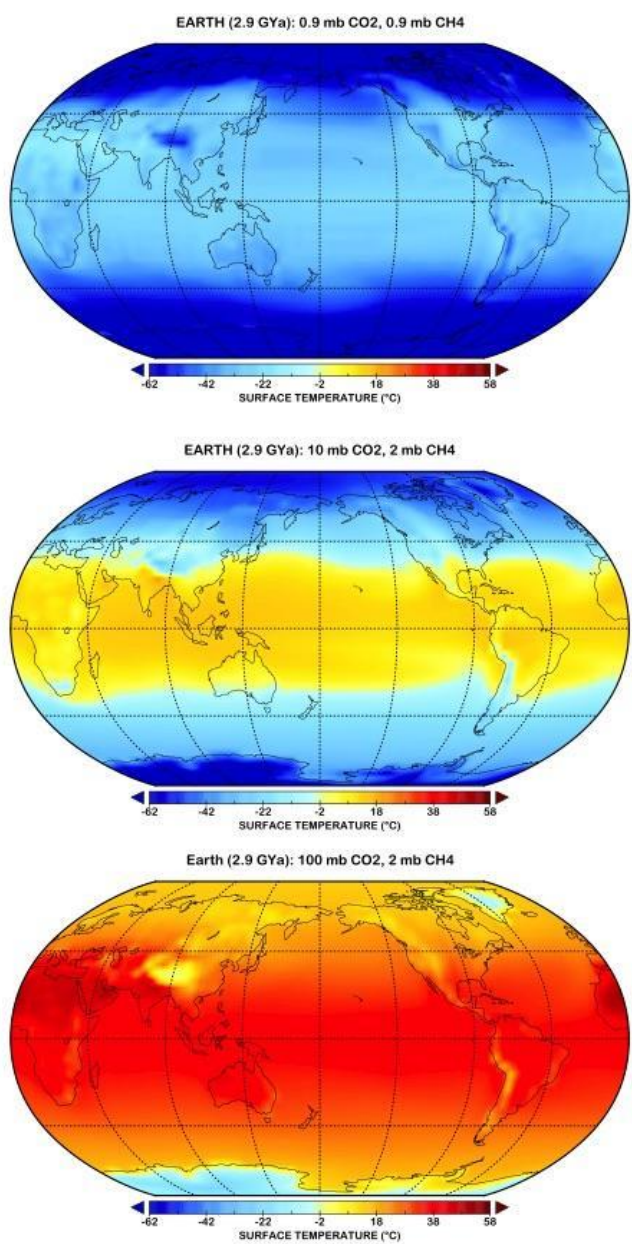

Figure 5.  2.9 Gya Archean Earth surface temperature simulated by the ROCKE-3D GCM (*Way et al.*, 2017) assuming atmospheric compositions for Cases (upper) A, (middle) B, and (lower) C of *Charnay et al.* (2013) and the solar spectrum from *Claire et al.* (2012).  Uncertainties due to the unknown land-ocean distribution are a few degrees. Simulations courtesy of Michael Way.

Figure 5 shows the effect of uncertainty in Archean $CO_2$ on its predicted climate. With only 0.9 mb of $CO_2$ (upper panel), the entire surface is glaciated, even with $CH_4$, because of the positive sea ice-albedo feedback. In fact if $[CH_4]/[CO_2] > 0.2$, an organic haze would have formed and further cooled the surface (*Sagan and Chyba*, 1997; *Wolf and Toon*, 2010; *Arney et al.*, 2016). With 10 mb of $CO_2$ (middle panel), an above-freezing tropical "waterbelt" is sustained, consistent with geologic and oxygen isotope data implying glacial conditions at subtropical to middle latitudes (*de Wit and Furnes*, 2016). $CO_2$ levels as high as those inferred by *Kanzaki and Murakami* (2015) (lower panel) produce a hot Archean climate, warmer than modern Earth and with ice only on polar land masses (if there were any at that time; Section 2.5). Bond albedo changes magnify the effects of the increased $CO_2$ and $CH_4$ since the planet darkens as it warms. This is the net result of decreases in sea ice and thicker, brighter clouds.

If explaining a habitable Archean Earth under the faint young Sun is a challenge, what then to make of ancient Mars, which was illuminated < 40% as strongly as modern Earth (Fig. 1) yet almost surely had surface liquid water (see Chapter 7; *Ehlmann et al.*, 2016; *Wordsworth,* 2016a)? Clearly this requires larger greenhouse gas concentrations than are needed to explain Archean Earth's climate. Today's thin $CO_2$ Mars atmosphere is thought to be largely due to escape of an early thick $CO_2$-$H_2O$ atmosphere supplied by volcanism (*Jakosky and Phillips*, 2001; Section 4). How much of the early $CO_2$ was instead sequestered as carbonate beneath the surface is highly uncertain, though. The carbonate-silicate cycle feedback might support a thicker $CO_2$ atmosphere on a cool ancient Mars with moderate liquid water given a less efficient chemical weathering sink than that on modern Earth (Section 3), perhaps with cyclic behavior (*Batalha et al.*, 2016). Evidence for surface carbonates from weathering is less than expected (*Edwards and Ehlmann*, 2015), though carbonates may be more widespread than originally thought (*Wray et al.*, 2016).

The thickness of Mars' early atmosphere is thus poorly constrained. *Kurokawa et al.* (2018) estimate a lower limit of 0.5 bar at 4 Gya, although *Tian et al.* (2009) reach a different conclusion with different model assumptions. Higher pressures (up to ~1-2 bars) during the valley network-forming period of the late Noachian (~3.8-3.6 Gya) are possible (*Kite et al.,* 2014; *Hu et al.*, 2015). However a thicker early Mars $CO_2$ atmosphere is not enough. The maximum greenhouse effect of $CO_2$ occurs at ~8 bars (*Kasting et al.*, 1993); at higher pressure Rayleigh scattering increases the Bond albedo sufficiently to prevent further warming. Even at pressures this high, $CO_2$ and $H_2O$ alone do not produce above-freezing temperatures on early Mars (*Kasting*, 1991).

Consequently, understanding the ancient Mars climate requires other warming agents. For example, cold $CO_2$ atmospheres form $CO_2$ ice clouds. These were initially thought to oppose warming because condensation reduces the temperature lapse rate and thus weakens the $CO_2$ greenhouse effect (*Kasting*, 1991). On the other hand, downward scattering of thermal radiation by $CO_2$ clouds enhances warming (*Forget and Pierrehumbert*, 1997; *Mischna et al.*, 2000; *Colaprete and Toon*, 2003). *Kitzmann* (2016), though, finds that more accurate radiative calculations produce a much smaller warming than estimated by earlier studies. Furthermore, in 3-D model simulations, $CO_2$ clouds do not completely cover the planet, further reducing any warming impact they might have (*Forget et al.*, 2013; *Wordsworth et al.*, 2013).

Other greenhouse gases may provide the additional warming, e.g., $CH_4$ and $H_2$ (*Ramirez et al.,* 2014; *Lasue et al.*, 2015; *Chassefière et al.*, 2016; *Kite et al.*, 2017a; *Wordsworth et al.*, 2017).  New estimates of $CO_2$-$H_2$ and $CO_2$-$CH_4$ collisionally induced absorption (*Wordsworth et al.,* 2017) suggest that ~1% levels of each gas could lift an ancient Mars with > 0.5 bars of $CO_2$ above freezing for geologically brief periods that might explain the valley networks and lakes.  If so, then the question is: Did ancient Mars have a warm climate with a long-lived northern hemisphere ocean (*Ramirez and Craddock*, 2018), or a generally cold and icy climate without an ocean but with enough episodic warming due to obliquity variations, impacts, or volcanic events to occasionally melt ice at high elevations (*Wordsworth*, 2016a)?  Clearly, a faint young Sun paradox for Mars still exists.

Furthermore, it is not even certain that the thick early Mars atmosphere was purely $CO_2$. The model of *Zahnle et al.* (2008) predicts that $CO_2$ should be photochemically unstable on early Mars under the cold conditions produced by the faint young Sun.  Including the effect of O escape, their model predicts gradual reduction of the $CO_2$ atmosphere to CO.  Depending on the time scale for O escape, as well as other factors such as volcanic activity and the composition of impact material, the time scale for conversion might be as short as tens to hundreds of million years.  Thus it is possible that inferences of surface pressure on ancient Mars from surface features could be biased by the assumption of a $CO_2$ atmosphere, rather than a mixture of $CO_2$ and CO of unknown proportions.

To some extent we know more about ancient Mars' climate than we do about that of ancient Earth. Earth's surface has been completely re-worked by plate tectonics, volcanism and weathering of rocks (Sections 2, 3). Thus, we do not know exactly when and where continents emerged above the ocean surface; samples of rocks from the Archean or the earlier Hadean Eons are relatively rare; and what evidence exists may or may not reflect the climate at the location at which the evidence was found.  By comparison, Mars did not have plate tectonics (Section 2) and lost much of its atmosphere in its first billion years (Section 4), preserving its history of habitable conditions via its surface geomorphology (Section 2) and geochemistry (Section 3). These provide not only evidence that surface liquid water was present but regional features that constrain the atmospheric general circulation and regional water cycle (*Wordsworth*, 2016a).

Venus represents the opposite extreme, since its resurfacing period (Sections 2, 3) left little or no geomorphological evidence of its ancient climate.  Venus' current 92 bar $CO_2$ atmosphere and strong insolation produce a runaway greenhouse climate that makes directly exploring its surface a challenge (see Chapter 16).  Thus no direct evidence of past surface liquid water on Venus exists; we only know that today it is dry.  If water was delivered to the terrestrial planets stochastically by embryos from the outer Solar System, Venus' initial inventory may have been less than one Earth ocean (*Raymond et al.*, 2006).  After the hot magma ocean and thick $H_2O$ steam + $CO_2$ phase on Earth and Venus (Section 3), did their evolutions diverge immediately or much later?  *Hamano et al.* (2013) and *Lupu et al.* (2014) find that the incident solar flux at Earth allows a magma ocean to cool and solidify fast enough for a water ocean to form.  At Venus, though, the stronger solar flux produces slower cooling and more water loss to space. Thus, Venus may or may not have ever formed a liquid water ocean.

The deuterium/hydrogen (D/H) ratio provides indirect evidence that Venus may have once had liquid water, although it might also be explained by an extended magma ocean phase with no surface water (Section 3.1). D/H is several orders of magnitude higher on Venus than on Earth (*Donahue et al.*, 1982). If this is due to water isotope fractionation after photodissociation due to preferential escape of lighter H atoms, D/H implies a putative global water ocean depth of 4-525 m (*Donahue and Russell*, 1997; *Chassefiere et al.*, 2012). For even a weak or moderately active young Sun, *Lichtenegger et al.* (2016) estimate that enhanced XUV fluxes were strong enough for hydrogen to escape efficiently; D and H escape rates would thus have been similar during the early blow-off phase, implying that today's observed D/H provides only a lower limit to Venus' possible past water loss. If water was replenished from later impactors or volcanism, though, the observed D/H does not require an early surface ocean (*Grinspoon*, 1993).

*Lichtenegger et al.* (2016) also show that oxygen does not escape as efficiently as hydrogen, implying that considerable $O_2$ should have been left behind on Venus after its first 100 Myr. Similar inferences have been made for habitable zone exoplanets orbiting M stars (*Luger and Barnes*, 2015). However, Venus' atmosphere today has little $O_2$ (see Chapter 16). Thus, the missing oxygen must have been removed by interactions between the atmosphere and solid planet. Different possible scenarios are discussed by *Lammer et al.* (2018), who suggest that the lack of oxygen today is easiest to explain if a liquid water ocean never formed and oxygen was simply incorporated into Venus' magmatic crust and oxidized the mantle during its first 100 Myr.

*If* primitive Venus did cool enough for liquid water to form, the faint young Sun might have been an opportunity for it to be an early habitable planet depending on the evolution of its spin. The "conservative" cloud-free habitable zone (Chapter 20) predicts that a solar flux only slightly higher than modern Earth's would turn it into a "moist greenhouse" or "runaway greenhouse." *Yang et al.* (2014b) show that a planet that rotates slowly enough creates rising motion and thick dayside clouds that raise the Bond albedo and shield the planet. *Way et al.* (2016) estimate that a planet rotating at current Venus' 243 d sidereal period 2.9 Gya ago could have maintained surface temperatures similar to modern Earth despite an incident solar flux 1.4 times that received by modern Earth.

We do not know whether Venus always rotated slowly or evolved to its slow current value via the solar gravitational torque on Venus' atmospheric thermal tide (*Dobrovolskis and Ingersoll*, 1980; *Correia and Laskar*, 2003; *Leconte et al.,* 2015). *Yang et al.* (2014b) and *Way et al.* (2016) find that a rotation period of ~1-2 months is slow enough to produce the stabilizing dayside cloud deck required for a habitable climate. This is determined by the radiative relaxation time of the atmosphere, since radiative heating/cooling of the dayside/nightside must be strong enough to drive upward/downward atmospheric motion. Thus, several possible scenarios exist: (1) Venus never had a water ocean because it lost its hydrogen in its cooldown phase or initially rotated too rapidly to prevent an early runaway greenhouse; (2) Venus cooled rapidly enough to form an initial ocean that transformed within the first 100 Myr into a runaway greenhouse, leading to escape of its hydrogen and part of its oxygen, with the remaining oxygen being oxidized into the hot surface; (3) Venus cooled rapidly enough and rotated slowly early enough in its history, or had enough water delivered later, to form a transient ocean, thick clouds, and a temperate climate for a finite time period, eventually losing its oxygen to the surface after it evolved into something closer to its current state. (Oxygen loss to space from Venus due to the solar wind is thought to have been modest [*Kulikov et al.*, 2006], although it may be supplemented by processes such as ion acceleration by electric fields due to pressure gradients and gravitational ion-electron separation [see *Brain et al.*, 2016 for a review of relevant processes]).

### 5.2.2 From ancient to modern times

Earth, Mars, and Venus have not warmed monotonically since their original cool-down after formation. Because of its smaller size, much of Mars' secondary atmosphere has escaped over time (Section 4) so its climate is colder today than in its past despite the brightening Sun. At least Earth and Mars appear to have also fluctuated between warm and cold periods throughout their histories. These climate fluctuations are of two fundamentally different types.

On Earth the first type consists of occasional catastrophic cooling events that caused glacial conditions to extend equatorward, sometimes fully glaciating the planet (see Chapters 1, 17). The oldest such "snowball Earth" period occurred ~2.2 Gya; there may have been regional events as early as ~2.9 Gya (*Kopp et al.*, 2005). The most well studied snowball periods are the Sturtian (~715 Mya) and Marinoan (~650 Mya), although it is not clear whether Earth was globally glaciated at these times (*Kirschvink*, 1992; *Hoffman et al.*, 1998; *Hoffman and Schrag*, 2002) or had a tropical liquid "waterbelt" (e.g., *Sohl and Chandler*, 2007).

External climate forcing cannot explain the snowball events. The 2.2 Gya glaciation occurred shortly after the "Great Oxidation Event" (GOE), a sharp rise in $O_2$ now thought to be merely the most dramatic change in a long history of the rise of $O_2$ (*Lyons et al.*, 2014; *Olson et al.*, 2018; see also Chapters 1, 17). Perhaps the release of large amounts of $O_2$ into the atmosphere chemically destroyed $CH_4$ and reduced its greenhouse effect (see Chapter 17). If so, this sets limits on late Archean $CO_2$. Figure 5 shows that a 2.9 Gya Earth with 100 mb of $CO_2$ is far too warm to later glaciate the Earth by removing $CH_4$. A midrange scenario with ~10 mb of $CO_2$ gives a cold mean climate but with open tropical oceans, a situation more susceptible to full glaciation if $CH_4$ is destroyed. If the putative ~2.9 Gya event is related to one of the "whiffs" of $O_2$ prior to the GOE (*Planavsky et al.*, 2014), this would also argue for modest $CO_2$ levels.

An equally interesting question is how Earth recovered from these snowball periods. *Walker et al.* (1981) had previously suggested in general that if a decrease in insolation had ever glaciated the Earth, the precipitation weathering $CO_2$ sink component of the carbonate-silicate cycle feedback mechanism would largely cease, allowing $CO_2$ to gradually build up via volcanic activity. *Kirschvink* (1992) first suggested that the observed Marinoan glaciation, though not driven by insolation change, might have ended via a similar volcanic $CO_2$ buildup. It is not clear whether inferred $CO_2$ levels were high enough to deglaciate a full snowball; even small areas of open ocean might provide enough of a sink to limit $CO_2$ buildup (*LeHir et al.*, 2008). Dust has been proposed as one possible mechanism assisting snowball deglaciation (*Abbot and Halevy*, 2010; *Abbot and Pierrehumbert,* 2010; *LeHir et al.,* 2010). *Abbot et al.* (2012) and *Abbot* (2014) have suggested that naturally occurring low-altitude stratocumulus clouds would provide a net warming effect over bright sea ice by enhancing downward longwave radiative fluxes to the surface. This process regulates Arctic sea ice temperatures on modern Earth (*Stramler et al.*, 2011) and may reduce the $CO_2$ needed to emerge from a snowball state.

Earth has also been anomalously warm in its past, e.g., in the Cretaceous (145-66 Mya), Paleocene-Eocene Thermal Maximum (56 Mya), and early Eocene (50 Mya) (*Lunt et al.*, 2017). These may have been caused by high fluxes of $CO_2$ and/or $CH_4$ from the ocean or interior, e.g., due to enhanced volcanism or destabilization of seafloor methane clathrates. These were *equable* climates, when the equator-pole temperature gradient almost vanished and tropical conditions existed at the poles. *Heller and Armstrong* (2014) argue that Earth is not an optimal habitable planet, and that "superhabitable" planets with more extensive, easily detectable biospheres may exist. Equable climate periods in Earth's past may be a prototype for superhabitable exoplanets.

The second, more modest type of climate fluctuation in Earth's geologic record is shorter term (~$10_4$-$10_5$ yr) advances and retreats of the ice sheets, such as the Last Glacial Maximum or "Ice Age" (~20Kya). *Milankovitch* (1941) calculated that Earth's orbital eccentricity, obliquity, and precession would vary with periods of ~100,000, 41,000, and 23,000 yr, respectively due to gravitational interactions with other Solar System objects. *Hays et al.* (1976) showed that temperatures inferred from deep-sea sediment cores were consistent with the Milankovitch cycles. Orbital variations do not primarily change total insolation, but rather its seasonal and latitudinal distribution. This causes sea-ice albedo and water vapor feedbacks and ocean-atmosphere exchanges of $CO_2$ that amplify the forced climate change (*Hansen et al.*, 2008). Milankovitch cycles still affect Earth's climate but are now dwarfed by anthropogenic greenhouse warming. For some rocky exoplanets, obliquity excursions over time might be much larger, with significant ramifications for their climates and thus for the location of the outer edge of the habitable zone (*Deitrick et al.,* 2018a,b; *Kilic et al.*, 2017, 2018; *Colose et al.*, 2019).

The valley networks and lakes that document liquid water and clement temperatures on Mars were created during the mid-late Noachian (~3.9-3.7 Gya), after a strong early escape period ~4 Gya (*Tian et al.,* 2009) had ended and allowed Mars' $CO_2$ atmosphere to remain fairly thick (*Irwin et al.*, 2005; *Fassett and Head*, 2008, 2011) until the Noachian-Hesperian boundary (~3.6 Gya). During this time, $CO_2$ may have been photochemically unstable and the atmosphere may have become enriched in CO (*Zahnle et al.*, 2008). The relative timing of important events is uncertain, but they include the cessation of Mars' magnetic field in the early Noachian (~4.1 Gya), subsequent atmospheric loss due to impact erosion, sputtering, and sequestration (*Brain and Jakosky*, 1998; Section 4), and the eventual collapse of the cooling Martian atmosphere as the $CO_2$ ice caps began to grow (*Jakosky and Phillips*, 2001; *Soto et al.,* 2015; *Kite et al.*, 2017a). After this the Mars climate was cold and arid for much of the Hesperian period (~3.4-3.1 Gya) and unfavorable for life, similar to present-day Mars.

Despite this, the evolution of Mars' post-Noachian climate was not monotonic. Alluvial fan deposits indicate that from the late Hesperian to early Amazonian (~3.4-2.8 Gya), wetter, habitable conditions occurred intermittently for ~$10_8$ years (*Kite et al.,* 2017b). Even in the mostly cold, hyperarid, uninhabitable late Amazonian period evidence exists for occasional liquid water on the surface that implies climate changes (*Morgan et al.*, 2011).

Like Earth, Mars exhibits Milankovitch cycles. Earth's obliquity variation is limited by its large Moon (*Lissauer et al.*, 2011). Mars' obliquity varies chaotically and over a larger range (*Armstrong et al.*, 2004; *Laskar et al.*, 2004; *Brasser and Walsh*, 2011). At low obliquity, Mars' ice caps expand and its climate cools; at high obliquity, melting takes place, the ice caps recede, and the climate warms, due to the lower albedo of Mars with less ice and because sublimation of $CO_2$ increases the atmospheric pressure and greenhouse effect. $CO_2$ snow migrates to low latitudes when obliquity is large (*Haberle et al.*, 2003; *Forget et al.*, 2006; *Mischna et al.*, 2013), where it can melt episodically. Obliquity fluctuations may explain several features of Mars' past climate and habitability, e.g., the "icy highlands" scenario for Noachian valley network formation (*Wordsworth et al.*, 2015) and, in combination with proposed "methane bursts" due to $CH_4$-clathrate destabilization, sporadic early Amazonian lake formation (*Lasue et al.*, 2015; *Chassefière et al.*, 2016; *Kite et al.*, 2017a; *Wordsworth et al.*, 2017).

Since we have little information about Venus' past, little can be said about its evolution. If Venus cooled too slowly for a water ocean to develop, it was always uninhabitable. Major issues include understanding the surface-atmosphere and escape processes (Sections 3, 4) that kept Venus from accumulating $O_2$ (*Kasting and Pollack*, 1983; *Gillmann et al.*, 2009; *Hamano et al.*, 2013; *Lichtenegger et al.*, 2016), which also has implications for exoplanets near and inside the inner edge of the habitable zone (*Luger and Barnes*, 2015; *Harman et al.*, 2015); a possible transition from an early tectonic to a stagnant lid geophysical regime (*Lenardic et al.*, 2008, 2016b); and the timing of major volcanism and plains emplacement (Section 2). There is furthermore no guarantee that Venus' evolution was destined to diverge from Earth's. The coupled climate-tectonic system may exhibit bistability (*Lenardic et al.*, 2016a): Internal dynamical perturbations due to unknowable inhomogeneities, or transient external surface temperature anomalies like those Earth is known to have experienced through its history, might conceivably make the difference between an evolutionary path that leads to plate tectonics vs. one that maintains stagnant lid conditions, with different consequences for the subsequent buildup of atmospheric $CO_2$ and thus the climate. If so, then Venus' and Earth's eventual fate might not always be replicated by rocky exoplanets in other stellar systems irradiated at similar levels.

If instead Venus formed a primitive ocean, are there plausible scenarios for an extended habitable period before the transition to today's thick $CO_2$ atmosphere and hot climate? A wet, slowly rotating early Venus could have remained habitable even to the present because of cloud shielding (*Way et al.*, 2018). This did not occur, so an ocean at the low end of D/H inferences, coupled with elevated $CO_2$ as for Archean Earth, might explain why. In this case the ocean evaporates and $H_2O$ is transported to the stratosphere (where photolysis and escape set in) more quickly. This leaves a dry atmosphere and subsequent warming and $CO_2$ buildup, as weathering ceases, surface carbonates decompose, the mantle partially melts and outgasses, and perhaps the interior changes its tectonic regime, producing its modern state (*Noack et al.*, 2012; *Lenardic et al.*, 2016b; Section 2, 3.2).

# 6. CONCLUSION

The inner Solar System appears to have had a rich history of habitability. At least one, most probably two, and possibly three planets were habitable soon after they formed and cooled down. Habitability did not evolve in a straight line but with fascinating ups and downs that are possible in a climate system that couples an atmosphere to an ocean, a land surface, a dynamic interior, a dynamic heliosphere, and a star that increases in brightness. The future looks to be less interesting, though. Venus' habitable period, if it existed at all, is long gone and will not return.

Mars' habitable era has also either come to a close or is nearing that stage. Even Earth only has ~1-2 Gy left before the Sun's luminosity increases by ~20% and it enters a runaway greenhouse. By then it may already have long lost its ocean, mostly through transport to the stratosphere, photolysis, and hydrogen escape (*Wolf and Toon*, 2015) and secondarily via subduction into the mantle (*Bounama et al.,* 2001). Of course the immediate problem for the inhabitants of Earth is the lifetime of the intelligent species that reads book chapters such as this one (*Frank and Sullivan*, 2014).

Taken together, Earth, Mars, and Venus provide an invaluable, but incomplete, template for thinking about the range of conditions under which planets can be habitable. Mars is a case study of the narrow window of habitability of small-mass planets. It forces us to confront our fairly primitive understanding of the factors controlling habitability, as we struggle to explain what we can see through the eyes of spacecraft instruments. Mars provides a good lesson about our limited imaginations as we begin to anticipate the possibilities on hundreds of planets in other stellar systems that will only yield their secrets slowly.

On the other hand, Mars is a missed opportunity as well. What if it had formed with a mass comparable to Earth's? Would a weakly illuminated planet with surface liquid water, a magnetic field, and a thick $CO_2$ atmosphere, more of which is retained rather than lost, have been continuously habitable over 4.5 Gy? As we attempt decades from now to interpret potential exoplanet biosignatures and ask what the probability of life is given a habitable planet (e.g., *Catling et al.*, 2018), how might our thinking have changed if 2 rather than 1 of two habitable planets in our Solar System had developed and maintained life? How much farther along would our understanding be if 1 bar $N_2$ atmospheres with trace amounts of $CO_2$ and $CH_4$ were not our only prototype for life-bearing planets?

In fact, a 1 bar atmosphere is not even indicative of our own planet's evolution: Earth may have seen large changes in $N_2$ and total pressure in past eons, though whether pressure was greater or less in the distant past is debated (*Marty et al.,* 2013; *Johnson and Goldblatt,* 2015, 2018; *Som et al.*, 2016; *Stüeken et al.*, 2016; *Zerkle and Mikhail*, 2017). These changes are due to both biotic and abiotic mechanisms (*Wordsworth,* 2016b), emphasizing the fact that the habitability and inhabitance of a planet are intertwined (*Goldblatt*, 2016). The clearest example of this is the unintended experiment that humanity is currently conducting on the habitability of our own planet - so much so that we are now considered to be in a new geologic epoch, the Anthropocene (*Steffen et al.,* 2007).

Venus is an equally valuable example for our exoplanet thinking, though in different ways. Given our Earth-centric view, could we have imagined the runaway greenhouse end state of evolution for our so-called "sister" planet if we had not seen an example? Before we knew about Venus' crushing pressure and harsh surface conditions, astronomers sometimes imagined it to be an Earth-like planet with possibly abundant water and ice clouds producing its high albedo (e.g., *Bottema et al.,* 1965). Now that we have discovered many exoplanets in the "Venus zone" located inside the inner edge of the traditional habitable zone (*Kane et al.*, 2014), might we find a few to be habitable if we do not rule them out without making the effort to characterize them?

A key feature of Earth's evolution is that it has remained inhabited since life first appeared. During periods when its global mean temperature was below freezing, it may have retained equatorial regions of liquid surface water. During more extreme snowball periods when it was fully glaciated, life still persisted in oceanic niches. Thus the simplistic idea of a habitable exoplanet sustaining surface liquid water is invalidated by our own planet's history (Chapters 1-5). Such cold, but habitable and inhabited, periods in our history may be of less practical interest in the search for life on exoplanets, because of the problems they pose for biosignature detection (*Reinhard et al.*, 2017; see also Chapter 17). Nonetheless, the inner Solar System's lessons about the temporal evolution of habitability provide needed perspective for our thinking about other stellar systems: The parched desert planet we find with a future telescope may have teemed with life eons ago, while a frozen, apparently lifeless planet we detect may blossom a billion years into the future, just as our time as a habitable planet is coming to an end.

**ACKNOWLEDGEMENTS.** We thank Helmut Lammer, Robin Wordsworth, and Vikki Meadows for insightful comments that helped us improve the manuscript. This chapter was made possible by collaborations supported by the NASA Astrobiology Program through the Nexus for Exoplanet System Science; by the Sellers Exoplanet Environments Collaboration; by the NASA Planetary Atmospheres Program; and by the NASA Psyche mission. This work was also funded by the Deutsche Forschungsgemeinschaft (SFB-TRR 170, subproject C06). This is TRR 170 Publication No. 67

## TABLES

TABLE 1: Contemporary loss mechanisms and their relevance for the escape of different upper atmospheric species at each of the three main terrestrial planets today.

|        | Thermal | Photochemical | Ion     | Sputtering |
|--------|---------|---------------|---------|------------|
| Venus  | ---     | ---           | O, C, H? | Ar?        |
| Earth  | H       | ---           | O, H    | ---        |
| Mars   | H       | O, N, C       | O, C, H? | Ar?        |